# Optimizing and Accelerating Space-Time Ripley's K Function based on Apache Spark for Distributed Spatiotemporal Point Pattern Analysis

**Yuan Wang[1,3], Zhipeng Gui[2,3,*], Huayi Wu[1,3], Dehua Peng[2,3], Jinghang Wu[2,3], Zousen Cui[1,3]**

1. The State Key Laboratory of Information Engineering in Surveying, Mapping and Remote Sensing, Wuhan University, Wuhan 430079, China;
2. School of Remote Sensing and Information Engineering, Wuhan University, Wuhan 430079, China;
3. Collaborative Innovation Center of Geospatial Technology, Wuhan University, Wuhan 430079, China;
* Correspondence: zhipeng.gui@whu.edu.cn (Z.G.); Tel.: +86-027-6877-7167 (Z.G.)

**Highlights:**

- Distributed space-time Ripley's K function optimized by spatiotemporal principles.
- Performance analysis of different optimization strategies.
- Guidelines for accelerating spatiotemporal analysis with HPC technologies.
- Visual analytics framework for analyzing point pattern of large POI datasets.
- Support various applications, such as industrial spatial agglomeration exploration.

**Abstract:** With increasing point of interest (POI) datasets available with fine-grained spatial and temporal attributes, space-time Ripley's K function has been regarded as a powerful approach to analyze spatiotemporal point process. However, space-time Ripley's K function is computationally intensive for point-wise distance comparisons, edge correction and simulations for significance testing. Parallel computing technologies like OpenMP, MPI and CUDA have been leveraged to accelerate the K function, and related experiments have demonstrated the substantial acceleration. Nevertheless, previous works haven't extended optimization of Ripley's K function from space dimension to space-time dimension. Without sophisticated spatiotemporal query and partitioning mechanisms, extra computational overhead can be problematic. Meanwhile, these researches were limited by the restricted scalability and relative expensive programming cost of parallel frameworks and impeded their applications for large POI dataset and Ripley's K function variations. This paper presents a distributed computing method to accelerate space-time Ripley's K function upon state-of-the-art distributed computing framework Apache Spark, and four strategies are adopted to simplify calculation procedures and accelerate distributed computing respectively: 1) spatiotemporal index based on R-tree is utilized to retrieve potential spatiotemporally neighboring points with less distance comparison; 2) spatiotemporal edge correction weights are reused by 2-tier cache to reduce repetitive computation in $L$ value estimation and simulations; 3) spatiotemporal partitioning using KDB-tree is adopted to decrease ghost buffer redundancy in partitions and support near-balanced distributed processing; 4) customized serialization with compact representations of spatiotemporal objects and indexes is developed to lower the cost of data transmission. Based on the optimized method, a web-based visual analytics framework prototype has been developed. Experiments prove the feasibility and time efficiency of the proposed method, and also demonstrate its value on promoting applications of space-time Ripley's K function in ecology, geography, sociology, economics, urban transportation and other fields.

# 1. Introduction

Ripley's K function is a multi-distance point pattern analysis method for studying the spatial arrangement or spatiotemporal distribution characteristics of geographic points in spatial analysis [1–3]. Typically, the pattern of points can be classified as randomness, clustering or dispersion by using density-based or distance-based measurements [2], such as quadrat analysis and nearest neighbor index. Effective approaches for detecting and analyzing these point patterns would be helpful to investigate and interpret the spatiotemporal point process hidden behind geographic phenomenon or social events. Among the approaches of point pattern analysis, Ripley's K function stands out in three aspects: (1) it is a distance-based and scale-independent method, so Modifiable Areal Unit Problem (MAUP) can be avoided; (2) its parameters can be derived from research area, not like the bandwidth in kernel density estimation that usually relies on experience [4]; (3) it considers not only the nearest neighbor like Nearest-Neighbor-Index, but also the other neighbors within the maximum distance, hence the information behind the point pairs can be fully utilized. Therefore, Ripley's K function has been widely applied in many fields, such as ecology [5], archaeology [6], epidemiology [7], criminology [8], sociology [9,10], economics [11–13] and , biology and medical science [14].

Point pattern analysis is critical for various scientific and commercial applications, but also incur computational challenges on spatiotemporal big data [15]. The emerging ubiquitous sensors and network technologies have brought us spatiotemporal big data. For example, plenty of location-aware volunteered geographic information [16] is generated and shared through social media platform such as OpenStreetMap, Twitter, Facebook, YouTube, Flicker and so on. Meanwhile, many location-based service providers embraced open data policy, and the taxi trip record in New York City provided by NYC TLC[1] is one of the representatives. Most of these data can be regarded as point events, such that the spatiotemporal pattern of these events such as industrial spatial agglomeration could be recognized through point pattern analysis methods, e.g., space-time Ripley's K function. However, Ripley's K function is compute-intensive and become extremely time-consuming when data volume increases for several reasons: (1) Ripley's K function measures point pattern through point pair distances, and time complexity of pairwise comparison between all points is quadratic; (2) Ripley's K function involves weight computation for point pairs to correct edge effect, which is positively correlated to the complexity of the geometric boundary of research area; (3) Ripley's K function requires a fair amount of simulations to support confidence evaluation for significance level of point pattern, which usually cost far more time than computation for observed points. The expected time cost of Ripley's K function would be even higher when extended from spatial dimension to spatiotemporal dimension. There have been desktop-based software packages that provide Ripley's K function and its extensions (e.g., Spatstat [17], Splancs [18], Stpp [19] in R), but the time efficiency is far from satisfying for large data volume, which affects the user experience of geoprocessing significantly [20] and impedes its further application. Hence, the optimization and acceleration of space-time Ripley's K function is urgent to enable efficient spatiotemporal point pattern analysis for big point datasets.

High Performance Computing (HPC) technologies have been applied to tackle the compute-intensive challenges brought by spatial analysis [21], whereas few studies concentrated on leveraging HPC technologies to Ripley's K function. Parallel and distributed computing frameworks, e.g., OpenMP, MPI, CUDA and MapReduce, were adopted to make use of computing power from multi-core CPU to massive GPU, and thus improve the performance of spatial analysis algorithms. In this context, multi-CPU-based [22] and massive-GPU-based [23] methods have been developed to accelerate Ripley's K function for large point datasets. Although great achievements have been made to accelerate the computing process of Ripley's K function by considering spatial distribution of points in task decomposition and weight reuse, these implementations are limited in scalability and workflow optimization due to the architecture and relative expensive programming cost of the parallel frameworks. Without framework-level supports and abundant third-party libraries, advanced spatial extensions, fault-tolerance mechanism and spatiotemporal-aware scheduling are hard to be achieved. Distributed frameworks like Apache Hadoop and Spark are gaining ground in big geospatial analytics [24,25], and Spark is getting prevalent due to its in-memory architecture and unified APIs for query and manipulating various data structures. However, existing parallel optimization methods of Ripley's K function couldn't fit well in such a distributed data pipeline, and a systematic distributed optimization method for Ripley's K function upon Spark is highly desired. Meanwhile, existing distributed systems for geospatial data analytics mainly focus on spatial dimension, and temporal dimension is seldom involved. These issues impede the development of potential applications of Ripley's K function and its variations for large POI dataset.

---

[1] https://www1.nyc.gov/site/tlc/about/tlc-trip-record-data.page

To address the issues above, this study improves the computing procedure of space-time Ripley's K in a distributed environment. Specifically, 1) spatiotemporal-index-based point pair acquisition is developed to avoid unnecessary pair-wise comparison; 2) 2-tier weight cache is designed to reuse spatiotemporal weights to decrease the cost from repetitive weight calculations; 3) spatiotemporal partitioning is utilized to reduce data redundancy among partitions in distributed systems; 4) customized serialization for spatiotemporal objects and indexes is leveraged to lower the cost of data transmission between nodes in the cluster. The performance experiments and a use case by using enterprise POI data in Hubei Province of China demonstrate the feasibility and efficiency of the proposed method. The main contributions of this paper are as follows:

- The employed spatiotemporal principles accelerate space-time Ripley's K function in distributed computing environments, and the parallel optimization study is extended from spatial dimension to space-time dimension.
- We analyzed the performance impacts and applicable scenarios of the proposed strategies, and developed a visual analytics framework prototype using Apache Spark, Web visualization APIs and geospatial toolkit, which provide guidelines for developing HPC-enabled spatiotemporal analysis algorithms.
- Application case demonstrates how distributed space-time Ripley's K function would support researches of space-time scale of industrial agglomeration by using big enterprise POI dataset.

The paper is organized as follows: Section 2 summarized related research about acceleration for Ripley's K function and spatial data processing in distributed computing framework. Section 3 introduced space-time Ripley's K function and analyzed its time complexity. Section 4 presented our method of distributed space-time Ripley's K function. Section 5 introduced the technical implementation of the proposed method as well as web graphical user interfaces (GUIs) of the developed prototype system. Experiments was analyzed and discussed in Section 6. Section 7 drew conclusions and discussed future research.

**2. Related work**

*2.1. Parallel Ripley's K function*

Ripley's K function has been highlighted for multi-scale spatial point pattern analysis in many natural science and socioeconomic fields [5–13]. It is an effective tool to identify the spatial distribution pattern of point events [2,3] and analyze its generation mechanism [5,6,8,11] by multi-distance measurement and hypothesis testing. Ripley's K function is powerful but compute-intensive since all point pairs are required to be traversed in estimation and simulations. As data volume increases, the time cost rises dramatically and even hinders its computability in standalone applications. To promote its application for big datasets, the acceleration methods of Ripley's K function have been investigated by optimizing its computing procedure and data parallelism using parallel computing frameworks.

The optimization of computing procedure is essential considering the high time complexity of original Ripley's K function workflow. Both estimation and simulation of Ripley's K function involve massive distance comparisons of point pairs within distance thresholds and weight computations to correct edge effects (refer to Section 3 and 4.3). Since not every point pair needs to be compared with the given distance threshold in nested traversals of points, a sort-based strategy was proposed to confine a rectangle for each point and avoid unnecessary distance calculations [22]. As different point pairs might have the same weights, a weights-reuse strategy was proposed to eliminate repetitive weight calculation in estimations and simulations of Ripley's K function [22]. Based on these optimization strategies, the time complexity of Ripley's K function could be effectively reduced.

The calculations of outer and inner traversals in estimation and simulation of Ripley's K function are independent with each other, which makes it feasible to parallelize Ripley's K function with multi-core CPUs and general-purpose GPUs. For a standalone computer, computation on outer traversal of each point could be dispatched to different threads of a multi-core CPU using OpenMP [22]. Each thread executes tasks of a subset of total points with access to the shared memory and maintain partial results in a local variable, then the partial results could be aggregated to obtain the final result of Ripley's K function. MPI was also investigated to parallelize Ripley's K function by using CPU resources from distributed nodes in a computer cluster [22]. In estimation phase, outer traversals are divided among distributed nodes, and are further parallelized with OpenMP on each node. Once all nodes complete computation, the partial results also need to be aggregated. In simulation phase, as each node keeps entire dataset in the MPI-based solution, simulations and corresponding calculations are executed independently on each node [22]. However, the data redundancy of MPI-based solution cannot be neglected for large scale datasets, and the performance of simulation phase will be limited by the hardware of each individual node. With the increasing popularity of the architecture of General-purpose Computing on Graphics Processing Units (GPGPU), GPUs have become a powerful option to solve large-scale data analysis and mining problems [26]. In this context, a CUDA-based scheme was proposed with parallel strategies, including variable-grained domain decomposition and thread-level synchronization [23]. Different from the CPU-based parallelization, in the GPU-based solution, outer traversals are handled by different CUDA blocks, while inner traversals are operated on different threads in each block simultaneously. Hence, synchronization within each block is designed to guarantee correct results of Ripley's K function. Related experiments have demonstrated the efficiency improvement of the parallel Ripley's K function comparing with that of the original sequential algorithm.

Although these parallel Ripley's K function solutions have effectively reduced time cost on large datasets, existing optimization methods only focused on spatial Ripley's K function. The computing procedure of space-time Ripley's K function is different from that of spatial one, so the corresponding optimization strategies should be redesigned. Meanwhile, HPC stack solutions still have limits comparing with big data stack solutions [27] (Table 1). (1) CUDA has relatively high hardware requirement, which might raise the threshold for its applications. Although other solutions in HPC stack can be adopted on commodity computers, OpenMP and Open MPI have limited scalability comparing with their counterparts in big data stack. (2) The fault-tolerant of HPC stack needs to be guaranteed by extra manual efforts from developers, while framework-level supervision and recovering mechanisms provided in big data stack can reduce development cost significantly and guarantee execution reliability. (3) Resource management for HPC stack is usually handled by developers with the support of additional third-party tools, while it is seamless integrated by distributed frameworks in big data stack. (4) HPC stack has relatively limited data sources and spatial extensions in general, while big data stack solutions provide built-in solutions for supporting heterogeneous data models and sophisticated data manipulation technologies. Besides, open-source spatial extensions on geoprocessing and spatial analysis available for big data solutions can benefit big data computation and develop Ripley's K function and its variations furtherly. Therefore, big data stack can provide solutions to optimize space-time Ripley's K function for supporting efficient, scalable and reliable spatiotemporal point pattern analysis.

**Table 1.** Comparison of HPC stack and big data stack for parallelism of Ripley's K function

| Software Stack | Framework | Language | Hardware Requirement | Data Source | Scalability | Fault-tolerance | Resource Manager |
|---|---|---|---|---|---|---|---|
| | OpenMP | C, C++ | Commodity | | Low | User-level | |
| HPC Stack | Open MPI | | Commodity | Network /Local File System | Medium | | Slurm, Torque |
| | CUDA | | High | | Medium | | |
| Big Data Stack | Hadoop | Java, Scala, Python, R | Commodity | HDFS, HBase, Hive | High | Framework-level | Yarn, Mesos, Spark Standalone |
| | Spark | | Commodity | | High | | |

*2.2. Distributed spatial data processing*

With more and more spatial data being generated and collected, a single computer would be insufficient to ingest, store and process them. In this context, distributed frameworks in big data stack have been applied to spatial big data management and computation.

Hadoop-based spatial systems, including Parallel SECONDO [28], Hadoop-GIS [29], SpatialHadoop [30], have been developed for spatial data processing. Parallel SECONDO integrates SECONDO, an extensible database system, to support spatial data. In this system, operations on spatial data are performed as distributed tasks which could be managed by Hadoop. However, it only supports uniform spatial data partitioning which makes it incapable of spatial data skewness problem. Hadoop-GIS implements SATO (a multi-strategy spatial data partitioning framework) and adopts local indexes to support efficient spatial query processing, which performs better for the skew in spatial data. SpatialHadoop provides powerful function to support more geometry types and various spatial partitioning techniques, e.g., uniform grids, R-Tree, Quad-Tree, KD-Tree, Hilbert curves. Benefiting from Hadoop software ecosystem, these solutions provide abundant fundamental spatial operators, which may be essential for developing advanced spatial analysis models, such as Ripley's K function. However, frequent I/O operations of Map/Reduce paradigm for Hadoop also lead to performance issues in iterative computation and limit its applications. As massive simulations based on the observed point dataset are required, Apache Spark would be a more appropriate choice to implement the optimized space-time Ripley's K function for its in-memory architecture, which can reduce intensive IO operations significantly compared with Hadoop.

Spark has drawn more attention in the field of GeoInformatics and Spark-based systems have arisen (Table 2), including SpatialSpark [31], GeoMesa [32], Magellan [33], Simba [34], GeoSpark [35]. SpatialSpark implements spatial range query and join query on the top of Spark Resilient Distributed Dataset (RDD), and it utilizes R-Tree partitioning and R-Tree index to speed up query. GeoMesa integrates with multiple databases and computing frameworks to enable large-scale spatial query. It establishes indexes based on space-filling curve (SFC), including z-order curve and Hilbert curve, to boost spatial query on Spark RDD. Magellan supports spatial range query and join query based on Spark SQL, and z-order curve is adopted to divide spatial objects. Simba also extends Spark SQL to implement operations for spatial data processing, and it provides R-Tree partitioning and R-Tree index to increase efficiency of queries. GeoSpark extends Spark RDD to represent complex geometrical shapes. It provides multiple spatial partitioning e.g., Uniform Grid, R-Tree, Quad-Tree and KDB-Tree, as well as local index including R-Tree and Quad-Tree to accelerate range query, K-Nearest Neighbor (KNN) query and range join query of large-scale spatial data. Besides tree indexes, Locality Sensitive Hashing (LSH) and graph indexes also have been applied to reduce complexity of KNN query [36]. These Spark-based systems provide spatial extensions to support spatial data models and fundamental spatial operations, however, there have been insufficient studies of spatiotemporal point pattern analysis. For example, SpatialSpark, GeoMesa and GeoSpark only support spatial objects and their manipulations, e.g., indexing, partitioning and query (Table 2). These spatial dimension features are insufficient to support space-time Ripley's K function; while spatiotemporal geometry models and additional operations such as edge-correction and generation of simulated spatiotemporal points are missing.

**Table 2.** Features of existing Spark-based spatial systems versus the requirements of space-time Ripley's K function

| Feature | Existing Spark-based Systems | | | | | Requirement |
|---|---|---|---|---|---|---|
| | SpatialSpark | GeoMesa | Magellan | Simba | GeoSpark | |
| Geometrical Objects | Spatial | Spatial | Spatial | Spatial | Spatial | Spatiotemporal |
| Indexing | R-Tree | SFCs | / | R-Tree | Multiple | Spatiotemporal indexing |
| Partitioning | R-Tree | / | SFC | R-Tree | Multiple | Spatiotemporal partitioning |
| Serialization | Default | Default | Default | Default | Customized (Spatial) | Customized (Spatiotemporal) |
| Specialized Features | Range query, Range Join | Range query, Range Join | Range query, Range Join | Range query, KNN query, Range Join | Range query, KNN query, Range Join | Edge Correction, Simulation, Weight Cache |

Therefore, we proposed our distributed space-time Ripley's K function implementation by developing a prototype system upon Spark and third-party spatial libraries directly. The optimization is achieved by leveraging spatiotemporal index, 2-tier weight cache, spatiotemporal partitioning and customized serialization. The proposed method can improve the procedure of space-time Ripley's K function and reduce computing cost from data skew, redundancy and transmission in distributed environment by utilizing spatiotemporal distribution characteristics, and making spatiotemporal point pattern analysis more applicable for big data scenario.

**3. Space-time Ripley's K Function for Spatiotemporal Point Pattern Analysis**

Space-time Ripley's K function is a statistical approach computed on space-time point events and estimates their second-order property based on point pair distance calculation [3]. It takes both number of points and distances between points into account, which allows for quantitatively evaluating how much the observed point pattern deviates from randomness at multiple spatiotemporal scales. The theoretical space-time Ripley's K function is calculated through dividing $E$, the expected number of points within spatial distance $s$ and temporal distance $t$, by the point intensity $\lambda_{st}$:

$$K(s,t) = \frac{E(s,t)}{\lambda_{st}} \quad (1)$$

Equation (1) characterizes the pattern of spatiotemporal points [37], where cylinder of base $\pi s^2$ and height $2t$ is centered on each point to count the number of neighbor points falling within. Then the total number of points $n$ is divided by the volume of the irregular cylinder formed by the study area and study duration, resulting in $\lambda_{st}$. It is expected that $K(s,t) = 2\pi s^2 t$ if the point distribution obeys complete spatiotemporal randomness (CSTR), $K(s,t) > 2\pi s^2 t$ if the points fit clustering within spatial distance $s$ and temporal distance $t$, and $K(s,t) < 2\pi s^2 t$ for dispersed space-time patterns. Using Equation (2), the space-time K function is formulated as:

$$\widehat{K}(s,t) = \frac{A \cdot D}{n^2} \sum_{i=1}^{n} \sum_{j \neq i} \frac{I_{s,t}(d_{ij}, u_{ij})}{\omega_{ij} v_{ij}} \quad (2)$$

where *A* denotes the area of study region and D is the duration of the study period. The product of *A* and *D* results in the volume of the irregular cylinder that is formed by the study area (base) and study period (height). $\omega_{ij}$ and $v_{ij}$ are spatial and temporal weighting function that corrects edge effects respectively. $d_{ij}$ is the spatial distance between the point $i$ and $j$, $u_{ij}$ is the temporal distance between point $i$ and $j$, $I_{s,t}(d_{ij}, u_{ij})$ is an indicator function defined in Equation (3):

$$I_{s,t}(d_{ij}, t_{ij}) = \begin{cases} 1, & (d_{ij} \leq s) \ and \ (u_{ij} \leq t) \\ 0, & otherwise \end{cases} \quad (3)$$

When the cylinder of a centered point exceeds the spatiotemporal scope of the study area, the expected number of neighbor points falling with the cylinder might be underestimated since there is no observed points outside the study area, and in turn generates so-called edge effect. Edge effects would bring biased result and impede effective analysis [17], and edge correction is what distinguish Ripley's K function from the various point pattern analysis methods. Alternative approaches to deal with spatiotemporal edge effects have been studied [38]. In this paper, we adopted widely-used isotropic edge correction method, where spatial edge correction weight is the proportion of circumference of the circle centered at point $i$ with radius $d_{ij}$ lying inside the study area, and temporal edge correction weight is 1 if both ends of the interval centered at point $i$ with length $2u_{ij}$ lie within study period or 1/2 otherwise [39]. Similar to the purely spatial Ripley's K function, space-time Ripley's K function could also be transformed to the space-time L function by Equation (4):

$$\hat{L}(s,t) = \sqrt{\hat{K}(s,t)/2\pi t} - s \quad (4)$$

where $\hat{L}(s,t) = 0$ under CSTR, $\hat{L}(s,t) > 0$ for clustered point patterns, and $\hat{L}(s,t) < 0$ for dispersed point patterns.

To test the statistical significance of the observed point pattern, the space-time Ripley's K function is evaluated for a large number of simulations. Monte Carlo, bootstrapping and random permutation are common methods for spatiotemporal point simulations [40]. The Monte Carlo method generates points in spatiotemporal space following CSTR; the bootstrapping method samples points randomly with replacement from the observed points; and random permutation generates a copy of the observed points and exchanges their temporal labels randomly. Based on the result of K values from simulated points, upper and lower envelope could be derived. If the observed K values is higher than the upper simulation envelope, spatiotemporal clustering for the corresponding spatial and temporal distances is statistically significant; while observed K values less than the lower simulation envelope indicate that point patterns exhibit significant spatiotemporal dispersion for the corresponding distances. $K(s,t)$ is then evaluated against multiple spatial and temporal distances to identify the scale ranges in which the point pattern follows a spatiotemporal random, clustered or dispersed pattern. Figure 1 is a flow chart that demonstrates the conventional computing procedure of spatiotemporal point pattern analysis using space-time Ripley's K function.

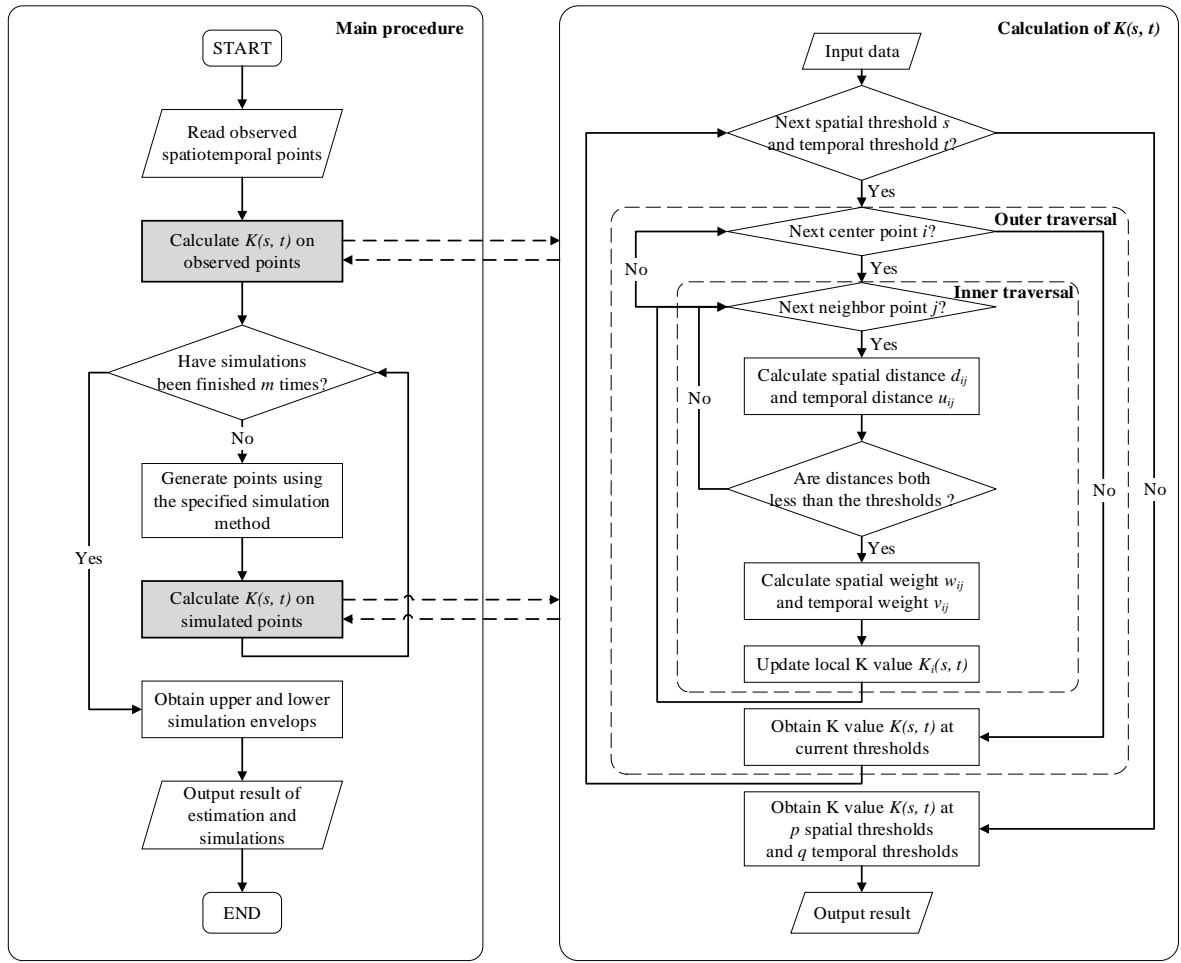

**Figure 1.** Flow chart of space-time Ripley's K function

For a case of spatial distance $s$ and temporal distance $t$, it takes $O(n^2)$ pair-wise comparisons to obtain $\hat{L}(s,t)$ for $n$ points. In each comparison, the edge correction weight is computed based on the vertexes of the spatial boundary and the duration of study period, which makes time complexity of $\hat{L}(s,t)$ become $O(v \cdot n^2)$ where $v$ denotes the number of vertexes. Given a series of spatial and temporal distances, $\hat{L}(s,t)$ is estimated with cost of $O(c_s \cdot c_t \cdot v \cdot n^2)$ where $c_s$ is the count of spatial distances, and $c_t$ is the count of temporal distances. For the simulated points, $\hat{L}(s,t)$ would be calculated by the same procedure above. Let $m$ represent number of simulations, the time complexity of spatiotemporal point pattern analysis using space-time Ripley's K function would be $O(m \cdot c_s \cdot c_t \cdot v \cdot n^2)$. In result, spatiotemporal point pattern analysis on study area composed of large dataset, complex geometry shapes at multiple spatial and temporal distances with sufficient simulations is extremely time-consuming theoretically. Therefore, space-time Ripley's K function is in urgent need of optimization and acceleration for spatiotemporal point pattern analysis on big point datasets.

## 4. Workflow and Strategies of Distributed Space-time Ripley's K Function

### 4.1. Workflow of distributed space-time Ripley's K function

To lower the barrier of spatiotemporal point analysis for large POI datasets, we proposed a distributed algorithm of space-time Ripley's K function in this section. Optimization strategies are designed to decrease the time complexity of the original algorithm, independently with the amount of computing resources to be utilized; while distributed processing approaches are adopted to improve the time efficiency furtherly with respect to the power offered by employing distributed computing resources. The following introduces the general idea of distributed space-time Ripley's K function.

Figure 2 shows how the distributed space-time Ripley's K function is operated over an Apache Spark environment. Tasks of the space-time Ripley's K function are divided according to the data storage of each worker node in the cluster, and partial results will be gathered on the master node. The job implementing the K function is characterized by a set of K functions parameters, including JAR package of algorithm implementation, point dataset identifier, study area, spatial and temporal distance threshold, edge correction method, simulation method, and number of simulations. Apache Spark parameters such as number of executors, CPU resources, and memory resources define the demand of computing resources for the job. The parameters above will be organized as a driver program. When the driver program is submitted to the master of cluster, appropriate resources for the job will be allocated; while calculation tasks will be generated according to the K function parameters. Built-in spatiotemporal partitioner in master node takes care of data partition among worker nodes. Customized serializer provides compact data representation of spatiotemporal objects for optimizing data transmission. Subsequently, the calculation of K function will be processed simultaneously by the workers (i.e., computing node) in cluster, which usually prefer to handle their local data partitions of the entire point dataset. The executors in the worker nodes are in charge of executing multiple threads that handle the calculation tasks assigned to them. In each executor, a spatiotemporal index builder is integrated for indexing local data partition. Meanwhile, spatiotemporal index for local data, hash-table-based weight cache and spatiotemporal objects, e.g., point, cylinder and envelop are cached in partitions of an executor for accelerating calculation. Once all the tasks are completed, the results will be transferred to and aggregated by the master node. Through this master-slave programming model in distributed computing framework, the distributed implementation of space-time Ripley's K function can be scheduled, monitored and accelerated.

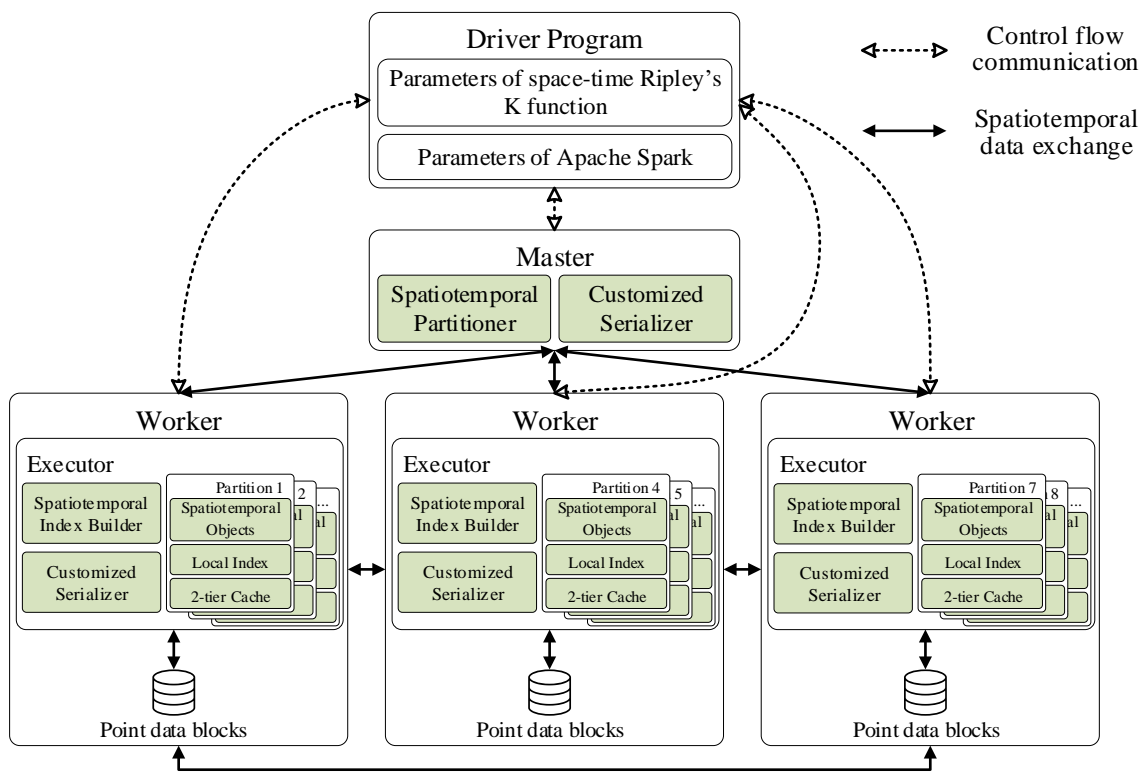

**Figure 2.** Master-slave programming model of distributed Space-time Ripley's K function in Apache Spark environment

The workflow of distributed space-time Ripley's K function is shown in Figure 3. Four strategies are adopted to optimize original space-time Ripley's K function: spatiotemporal-index-based point pair acquisition, 2-tier cache strategy for spatial and temporal weight reusage, spatiotemporal partitioning, and customized serialization for spatiotemporal objects and indexes. In the main procedure, both observed points and simulated points will be spatiotemporally partitioned before the calculation of K values to balance working load and reduce IO overhead among nodes in the cluster. Then spatiotemporal indexes will be built on each partition to boost the query for neighbor points, and 2-tier cache will be utilized to reuse repetitive spatiotemporal edge correction weights. In addition, customized serialization supports compact data transmission for spatiotemporal objects and indexes, across through entire distributed computing procedure. The design details of these four strategies will be explained in following sections.

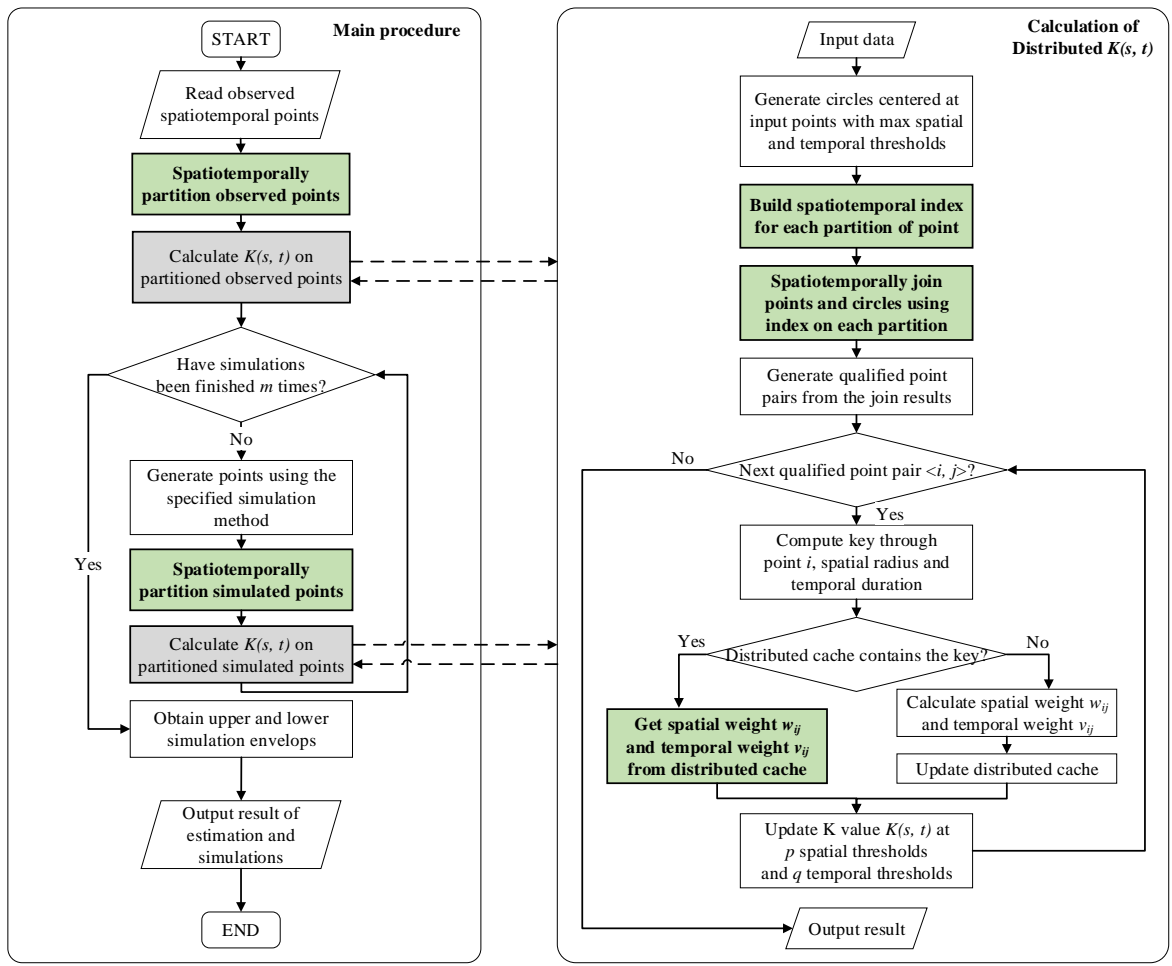

**Figure 3.** Workflow of optimized space-time Ripley's K function in a distributed computing environment

*4.2. R-tree-based spatiotemporal index for point pair acquisition*

The calculation of the space-time Ripley's K function requires nested traversals on the point data. The outer traversals must cover every point, and the inner traversals theoretically only need to find the neighboring points that lie within the spatial and temporal thresholds. In this paper, the point pair acquisition is regarded as a query task to avoid unnecessary traversals. The input of the query includes spatiotemporal point $p$, spatial threshold $s$ and temporal threshold $t$, and the goal is to find all the points in the cylinder centered on $p$ with base of $\pi s^2$ and height of $2t$. For such a query task, spatiotemporal index can quickly narrow the query scope, which decreases the comparison times of inner traversals.

The research on spatiotemporal index is mainly developed from spatial index. It can be divided into the following types: quadtree-based index, R-tree-based index, KD-tree-based index, and geohash-based index. Quadtree is extended to an octree in the spatiotemporal application [41,42].

The spatiotemporal cube of study area and duration is divided into octaves according to the max number of child nodes and tree-depth, such that finally a series of spatiotemporal sub-cubes of different levels are derived. The advantage of the octree lies on its simplicity and ease of implementation. However, the structure of the octree can become unbalanced for uneven distribution of spatiotemporal points, which means that the branches corresponding to dense areas will be deeper, and the branches corresponding to sparse areas will be shallower, resulting in unstable query efficiency of octree. As an extension of the B-tree in two-dimensional space, R-tree is a highly balanced tree [43]. Data objects are stored in the leaf nodes, and each non-leaf nodes only record the aggregated minimum bounding rectangle (MBR) from their children. In the spatiotemporal scenarios, the MBR becomes a spatiotemporal cube; in this respect, many variants have been proposed [44–48]. R-tree-based spatiotemporal index provides better query efficiency because of the balanced tree structure and moderate tree height, but the operations of insertion and deletion are complicated, and the dead space cause by overlap between MBRs of the tree nodes would lead to invalid query results. KD tree is the extension of binary space partitioning (BSP) tree in multidimensional space [49]. Each node contains one data object and represents a hyperplane of one dimension, as such the other data objects divided by the hyperplane could be found in the child nodes. For data objects in a spatiotemporal cube, KD tree will take turns to determine the hyperplane according to x direction, y direction and time direction. The structure of KD tree makes it advantageous in accurate query and KNN query; however, as each node has only two child nodes, the tree will be rather deep such that it doesn't perform well in range query. Geohash utilizes space-filling-curve (SFC) to map high-dimensional space to one-dimensional space, thus resulting in a hash string describing the high-dimensional space. For spatiotemporal data processing, geohash is usually spliced with timestamps to refer to the spatiotemporal cube in which the data object belongs [50]. In range query for a geohash-based index, the query scope will be converted into a hash string, such that data objects in spatiotemporal cubes having a hash string with the same prefix will be added to the result. However, neighbor cubes which have different hash strings might also contain data objects for the query, hence they need to be checked for a complete and correct query result. Although the regularity of geohash makes it appropriate for data storage and management in distributed systems, the abruptness of the SFC complicates the logic of query. As the performance of range query plays a significant role in point pair acquisition for space-time Ripley's K function, a R-tree-based spatiotemporal index will be a proper choice. In this paper, R-tree with Sort-Tile-Recursive (STR) algorithm [51] is selected to optimize procedure of inner traversals in space-time Ripley's K function as illustrated in Figure 4, where the points will be bulk loaded into the tree and overlap between the spatiotemporal MBR of nodes will be avoided.

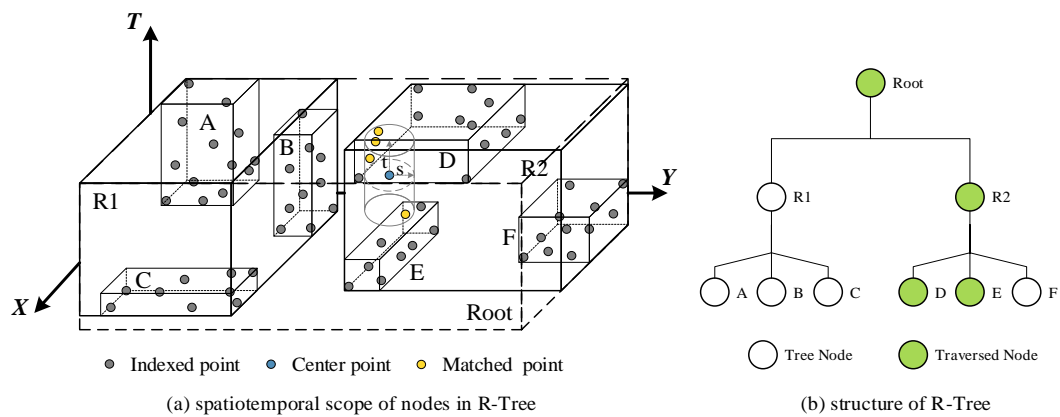

**Figure 4.** R-tree with Sort-Tile-Recursive (STR) algorithm for spatiotemporal points

Point pair acquisition based on the R-tree is designed to reduce the number of inner traversals. To achieve this goal, two steps are required. The first step is to build R-Tree for spatiotemporal points. According to the threshold for child nodes count, spatiotemporal points will be first equally divided along the x, y, and time directions in turn, and then a series of spatiotemporal cubes containing close amount of points will be derived. Then the spatiotemporal cubes will be recursively divided in the same way, until only one cube left, which become the root of R-tree. The second step is to query point pairs through the R-tree. A series of cylinders centered at the spatiotemporal points (blue point in Figure 4(a)) with radius that equals spatial threshold and height that equals double temporal threshold will be constructed as query scope. Then the query scope will be compared with spatiotemporal cubes of the tree nodes. If they intersect with each other (green nodes in Figure 4(b)), same operation will be performed on the child tree nodes, until the lead nodes are reached. For the data objects in leaf nodes, they will be directly compared with the query scope, and matched points (yellow points in Figure 4(a)) will be added to the result. Therefore, less comparisons are actually made only for those nodes which potentially match the respective query. Through this optimization strategy, the time complexity of the space-time K function can be optimized from $O(m \cdot c_s \cdot c_t \cdot v \cdot n^2)$ to $O(m \cdot c_s \cdot c_t \cdot v \cdot n \cdot \log n)$ if the built R-tree index is balanced.

*4.3. 2-tier cache strategy for spatial and temporal weight reusage*

Edge correction weight for a spatiotemporal point pair is computed by the product of spatial and temporal weights. When the study area becomes more complex, weight calculation will be time-consuming. However, the repetition of weight calculation occurred among the spatiotemporal point pairs can be eliminated, which mainly maps to following two cases. In the first case, there might be multiple neighbor points having the same spatial distance and temporal distance from the same center point, and their isotropic correction weights can be calculated as described in Section 3. Obviously, the value of spatial isotropic correction weight is determined by coordinate of the center point and spatial distance of the point pair, while the value of the temporal isotropic correction weight is determined by timestamp of the center point and temporal distance of the point pair. Therefore, the spatial weight and temporal weight might have the same value for the multiple spatiotemporal point pairs, which could be reused after the first calculation (Figure 5). In the second case, the coordinates and timestamps of simulated points fully or partially come from the estimated points, thus the values calculated in estimation could be reused in simulations. The degree to which the weight could be reused in simulations depends on the simulation methods. As for random permutation commonly used in spatiotemporal analysis, timestamps of the observed points would be randomly exchanged. It means that most calculations for spatial and temporal weights in simulations could find reference in estimation respectively since the two weights have independent calculation processes. It is worth noting that spatiotemporal weight calculations only happen if the point pair satisfies the requirement of both spatial threshold and temporal threshold. After random permutation, a small number of point pairs in estimation will disappear, and new point pairs might be generated. Therefore, partial of point pairs in simulations still need to calculate weights from scratch. In general, the occurrence of the first case is related to the spatiotemporal distribution of points, while the frequency of the second case will increase when the spatial and temporal threshold become larger.

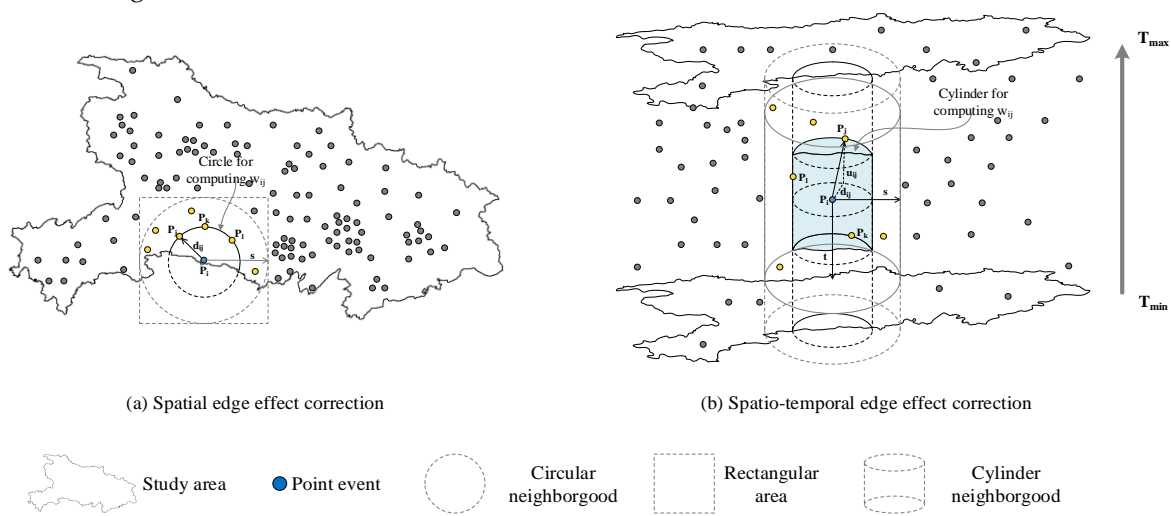

(a) Spatial edge effect correction     (b) Spatio-temporal edge effect correction

Study area   Point event   Circular neighborgood   Rectangular area   Cylinder neighborgood

**Figure 5.** An example of spatiotemporal isotropic edge correction

To avoid repetitions in the aforementioned two cases, spatial and temporal weights can be cached into two hash tables separately for reusing. Considering the application requirements of random permutation, cache structures for spatial and temporal weights are designed as independent 2-tier hash tables instead of being bound to observed points for maximizing the reusage in both estimation and simulation (Figure 6). Spatial weight cache and temporal weight cache are filled with $<p_s,<d,\omega>>$ and $<p_t,<u,v>>$ entries respectively, where $p_s$ and $p_t$ are coordinate and timestamp of the center point, $d$ and $u$ are the spatial distance and temporal distance, and $\omega$ and $v$ are the spatial weight and time weight. The key in the first-tier hash table is the hash value of $p_s$ and $p_t$ respectively, and the corresponding value is the second-tier table for $p_s$ and $p_t$. While, the key of the second-tier hash table is the hash values of $d$ and $u$ respectively, and the corresponding value is the spatial and temporal weight. Suppose we need to calculate the spatiotemporal weight of a point pair composed of point $p_i$ and neighbor point $p_j$, the spatial distance is derived as $d_{ij}$, the temporal distance is derived as $u_{ij}$, then they will be compared to existing $<p_s,<d,\omega>>$ and $<p_t,<u,v>>$ entries. If the keys can be found, the corresponding spatial weight $\omega$ and temporal weight $v$ will be directly taken out, otherwise the weights have to be calculated. During the estimation phase, the new $<p_s,<d_{ij},\omega>>$ and $<p_t,<u_{ij},v>>$ are inserted into the two hash tables. While, in the simulations, most weights could be obtained from the two hash tables, and a small number of new $<p_s,<d_{ij},\omega>>$ and $<p_t,<u_{ij},v>>$ will not be written into the two hash tables to reduce overhead for data consistency.

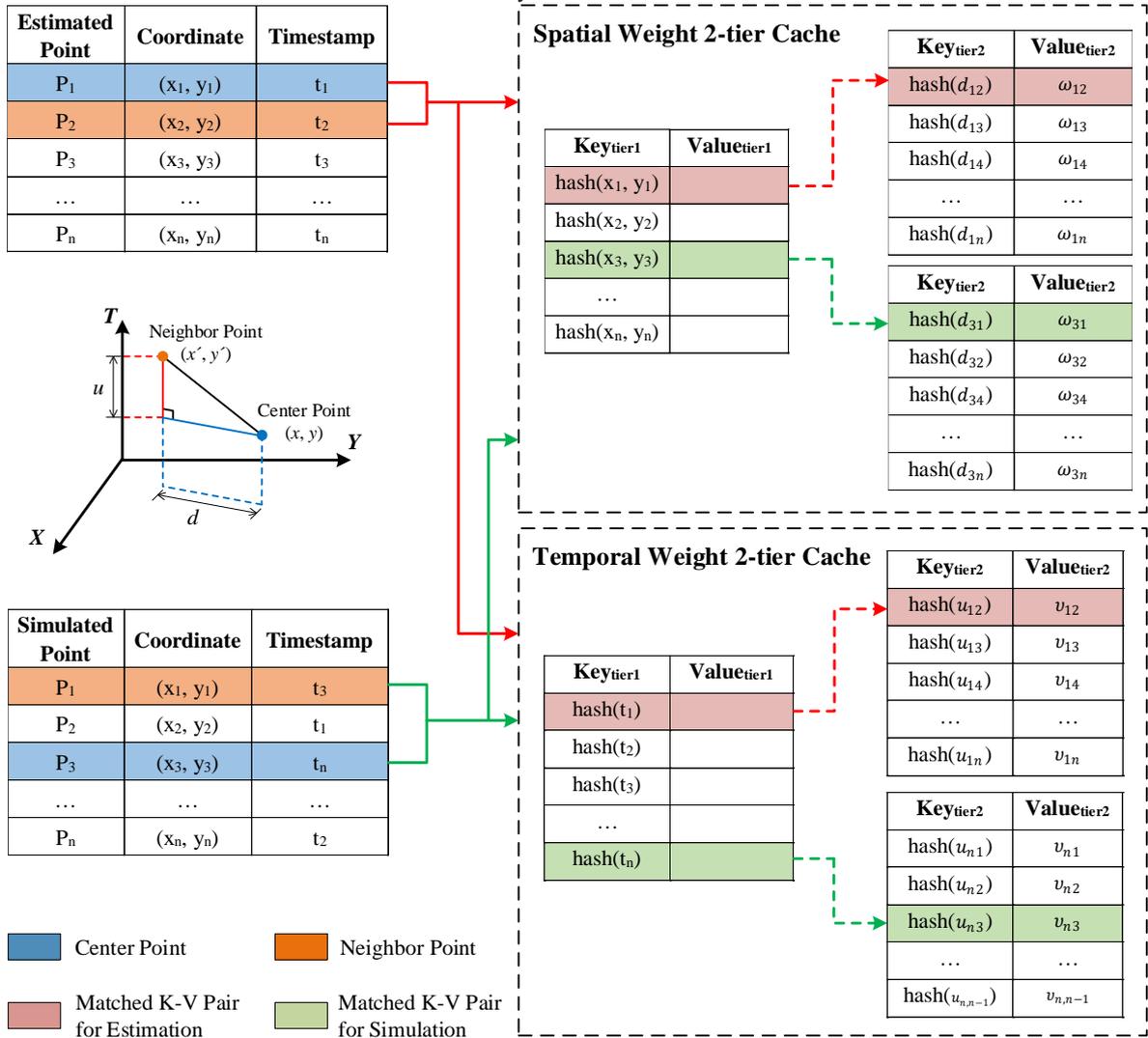

**Figure 6.** Spatial and temporal weight reusing with 2-tier cache for estimation and simulations

It is noteworthy that the spatial resolution of spatial coordinates and the time granularity of timestamps vary in applications. For example, in some datasets, there might be close spatial coordinates and spatial distances, but they are rarely exact the same value; while the timestamps might easily replicate due to the coarse time granularity. In this context, spatial coordinate tolerance and spatial distance tolerance are proposed to quantize spatial coordinates of the points and spatial distances of the point pairs. Apparently, smaller tolerance will bring higher precision, but larger tolerance will improve the effect of cache. Since the time complexity of insertion and lookup in a hash table is $O(1)$ [52], the time complexity of space-time K function will be further reduced from $O(m \cdot c_s \cdot c_t \cdot v \cdot n \cdot \log n)$ to $O(m \cdot c_s \cdot c_t \cdot n \cdot \log n)$ through 2-tier cache ideally.

*4.4. KDB-tree-based spatiotemporal partitioning for effective domain decomposition*

Moderate data redundancy among data partitions is essential for avoiding unnecessary IO cost on data transmission between computing nodes, and eventually accelerating the computation of space-time K function since the calculation process of each center point involves its spatiotemporal neighboring points within certain spatiotemporal distance threshold. A typical way is to establish a point buffer based on the spatiotemporal cube that contains the point dataset of each node, as well as the spatiotemporal neighboring points need to be read from other nodes, which is called "ghost buffer redundancy" in this study (Figure 7(a)). The size of the buffer is determined by the spatial threshold and the temporal threshold. Besides, the "cylinder intersection redundancy" proposed in this study (Figure 7(b)) provides another solution. As mentioned in the Section 4.2, the acquisition of the point pairs in space-time K functions can be regarded as a series of range queries. Therefore, a series of spatiotemporal cylinders can be generated based on the spatiotemporal points and spatiotemporal thresholds, and those who intersect the spatiotemporal cube of nodes in the cluster will be assigned to the corresponding node. The computing times of ghost buffer redundancy and the cylinder intersection redundancy are the same, which are the product of the number of computing nodes and the number of points. However, cylinder intersection redundancy has two advantages. Firstly, cylinder intersection redundancy can reduce data redundancy as points that do not meet the spatiotemporal threshold can be more accurately excluded. Secondly, the execution logic of cylinder intersection redundancy is simpler and easier to understand. The result of cylinder intersection redundancy includes the cylinders generated from both local points and buffer points, then the subsequent calculation can be completed directly based on the points and the cylinders. However, for ghost buffer redundancy, local points will only be involved in outer traversals, while buffer points must be included in the inner traversals, which makes the procedure more complicated.

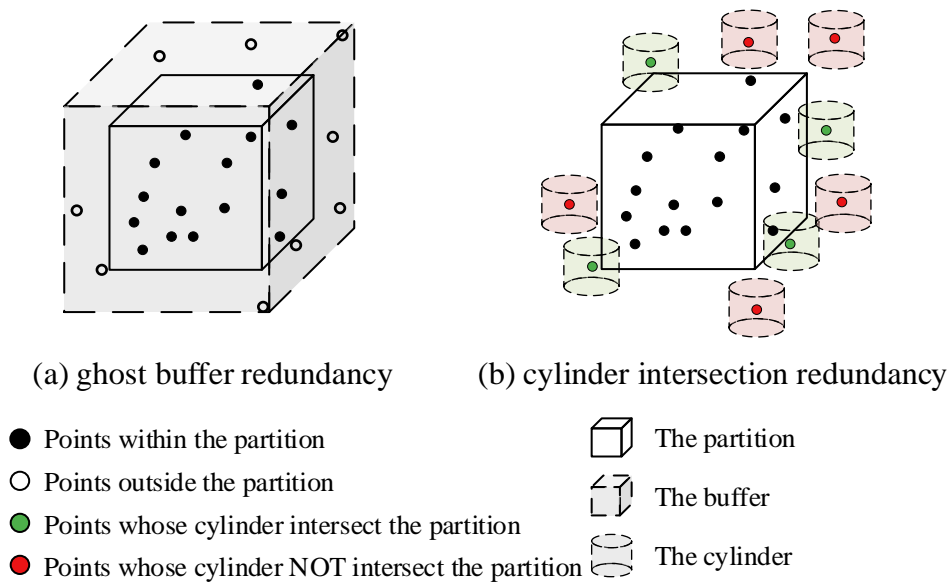

(a) ghost buffer redundancy     (b) cylinder intersection redundancy

- ● Points within the partition
- ○ Points outside the partition
- ● Points whose cylinder intersect the partition
- ● Points whose cylinder NOT intersect the partition

- ▢ The partition
- ▨ The buffer
- ⌭ The cylinder

**Figure 7.** Two kinds of data redundancy to guarantee correct result of space-time K function

Although cylinder intersection redundancy is more efficient on data volume, unnecessary redundancy still occurs as the spatiotemporal points are randomly stored in computing nodes. Thus the spatiotemporal cubes of nodes inevitably overlap with each other. As a result, the spatiotemporal cylinders may be divided to multiple nodes, but many of them are unnecessary (Figure 8(a)). This will cause a lot of invalid data transmission and memory occupation, affecting the computational efficiency of the distributed space-time K function implementation. In order to reduce the influence of data disorder, data partitioning must be optimized according to the law of space-time point distribution, so that spatiotemporally adjacent points will be divided into the same partition, and data redundancy can be further reduced (Figure 8(b)).

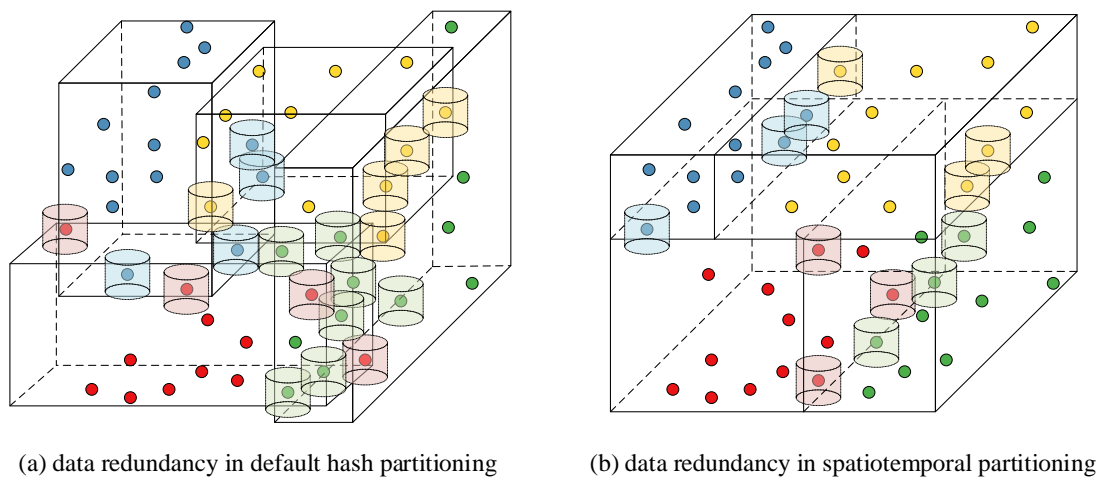

(a) data redundancy in default hash partitioning    (b) data redundancy in spatiotemporal partitioning

**Figure 8.** Data redundancy in different partitioning

Spatiotemporal indexes can be used in spatiotemporal partitioning, but the selection preference will be different with in point pair acquisition described in Section 4.2. Acquisition of point pair in the space-time K function focus on the efficiency of range query, while spatiotemporal partitioning pays more attention to the partition coverage and data balance. On the one hand, spatiotemporal objects must be covered by the spatiotemporal scopes of partitions. Otherwise, an extra partition is needed to store the points not covered by any spatiotemporal scopes of the partitions, which will destroy the spatiotemporal neighboring characteristics of the data in the partition. On the other hand, the number of points in the partitions should be as close as possible to avoid uneven workload of computing tasks on them. As can be seen from Figure 9: (1) Grid-based partitioning result would be able to cover the entire dataset, such as geohash, but the partitions derived are less in balance; some are quit dense and others sparse since the spatiotemporal distribution of points has its own patterns; (2) quadtree-based partitioning result can also cover the entire dataset, and dense partition will be divided into quarters, but sparse partitions still occur. (3) R-tree-based partitioning can divide the sample data evenly while maintaining the space-time proximity, but the entire dataset are not guaranteed to be fully-covered; (4) KDB-tree [53] partitioning generates partitions by dividing the study area instead of the MBR of data, so dataset will be fully covered. Meanwhile, KDB-tree enables the generation of near-balanced partitions. Therefore, this study uses KDB-tree based spatiotemporal partitioning to accelerate the calculation of space-time K function in a distributed environment.

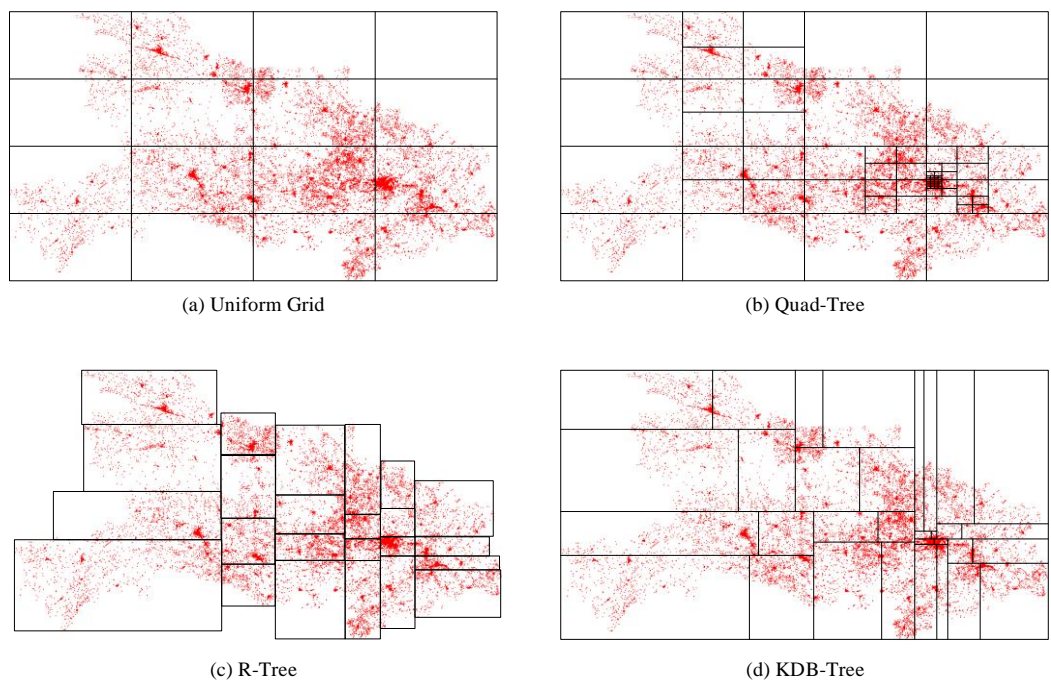

**Figure 9.** Data partitioning under different spatial indexes

In our method, spatiotemporal partitioning is accomplished by building spatiotemporal index using sample data to accelerate spatiotemporal scopes generation, which follows the next six steps (Figure 10). (1) spatiotemporal points are read from storage system; (2) the points are randomly sampled and sent to the master. According to relevant research, 1% of the data samples are sufficient to obtain high quality partitions [54]; (3) a KDB-tree-based spatiotemporal index is built on the sample points by the master, and the number of partitions is decided by the construction parameters for the tree, such as maximum number of child nodes, maximum number of items; (4) the index is delivered to the workers; (5) the workers query the leaf node to which each point belongs and constructs key-value pair $<id, p>$, where $id$ is the unique identifier for spatiotemporal envelop of the leaf node, and $p$ denotes the spatiotemporal point; (6) distributed points is repartitioned according to the key of pair, and key-value pairs with the same key is divided into the same partition. Eventually, spatiotemporally partitioned points can be derived from the value of pairs.

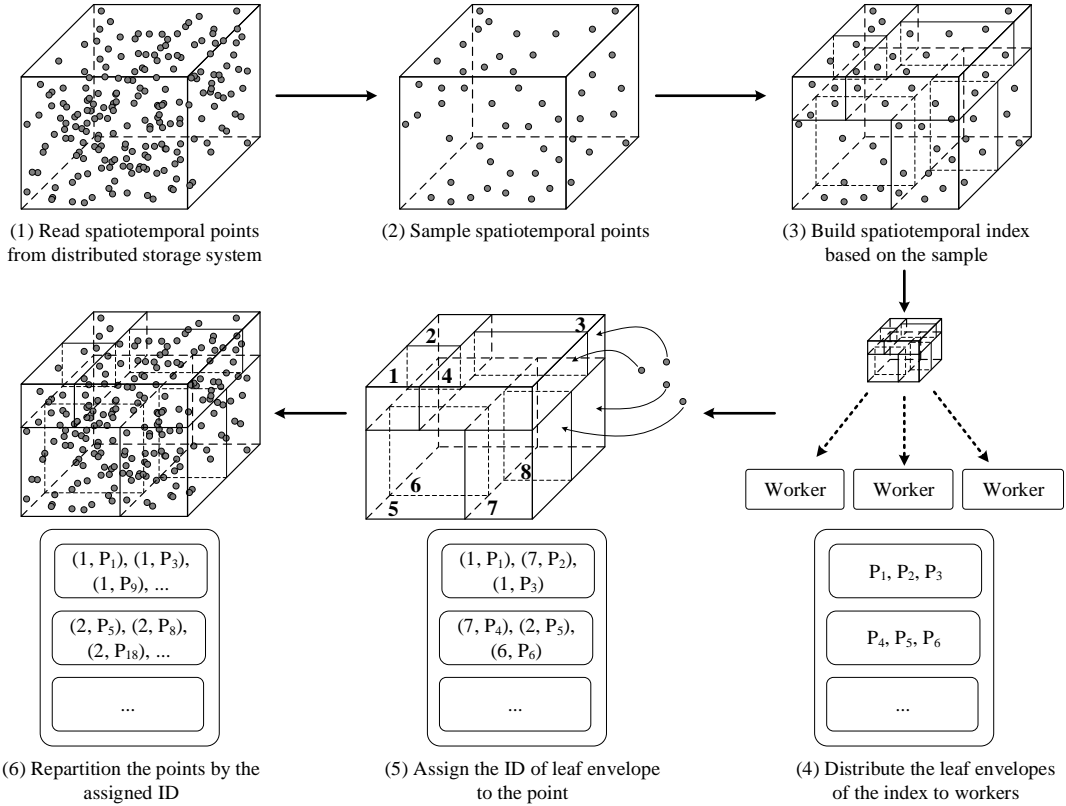

**Figure 10.** Process of spatiotemporal partitioning

*4.5. Customized serialization for spatiotemporal objects and indexes*

For the tasks of distributed space-time K function, data transmission is inevitable between different nodes of the cluster for exchanging spatiotemporal objects and spatiotemporal indexes. During transmission, the involved data structures are serialized by the sender machine into a byte array, and then the receiver machines deserialize the received data into memory in general. Big data processing frameworks, such as Spark, provide serializers with sufficient capabilities to handle simple data objects such as integer and double, but for more complex spatiotemporal objects, a compact representation cannot be achieved [55], which might cause more bytes generated and transferred in the cluster.

To solve this problem, this study designs a custom serialization for spatiotemporal objects including spatiotemporal points, spatiotemporal cylinders and spatiotemporal envelopes, as well as spatiotemporal indexes, i.e., KDB-tree and R-tree. The spatiotemporal cylinders need to be transmitted between different nodes during the generation of cylinder intersection redundancy, and the spatiotemporal points will be transmitted as the center of the spatiotemporal cylinders. KDB-tree needs to be serialized and deserialized when the master distributes the spatiotemporal partitioner to workers; while R-tree would be transferred between nodes for the scheduling of tasks. For example, when a job failed on one node, the data partitions and local indexes need to be transferred to another node for recalculation. The spatiotemporal envelope is used as representations of the spatiotemporal range for the tree nodes in spatiotemporal indexes.

As the JAR packages are distributed to all the nodes during the computation, the class description information of the spatiotemporal objects and indexes have already existed in the nodes, hence only the key attributes of object instances are needed to be serialized. The serialization and deserialization of spatiotemporal objects is performed according to the structure depicted in Figure 11. Apparently, the structure of byte array varies with the type of the spatiotemporal object being transmitted. For a spatiotemporal point, the spatial coordinates, the count of spatiotemporally overlapped points, start time, end time and zone ID will be serialized after its type ID, which indicates its object class type. For a spatiotemporal cylinder, besides type ID, the center of the cylinder is serialized as a spatiotemporal point, and spatial radius, temporal radius and temporal unit are added to the byte array sequentially. For a spatiotemporal envelope, the four values of coordinate, start time, end time, zone ID and envelope ID are serialized in the following of its type ID.

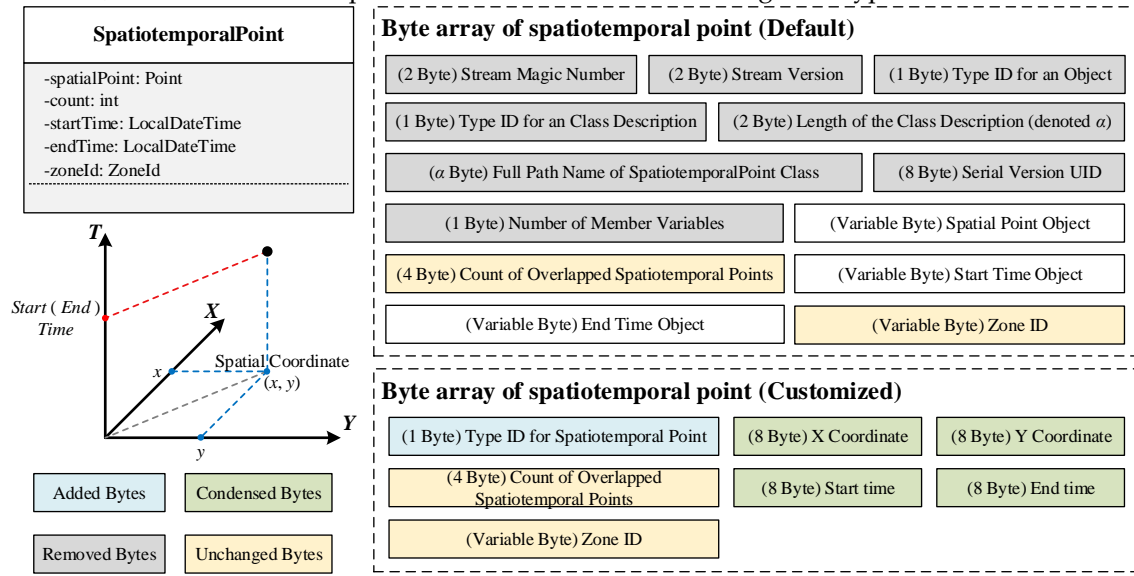

(a) Class of SpatiotemporalPoint and its byte array structure with default and customized serializer

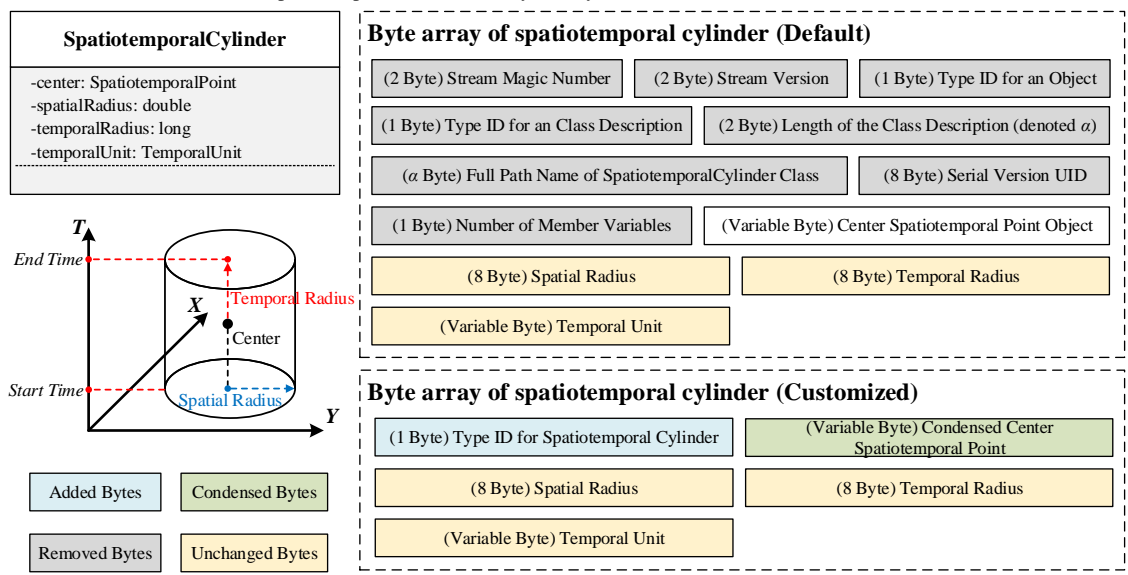

(b) Class of SpatiotemporalCylinder and its byte array structure with default and customized serializer

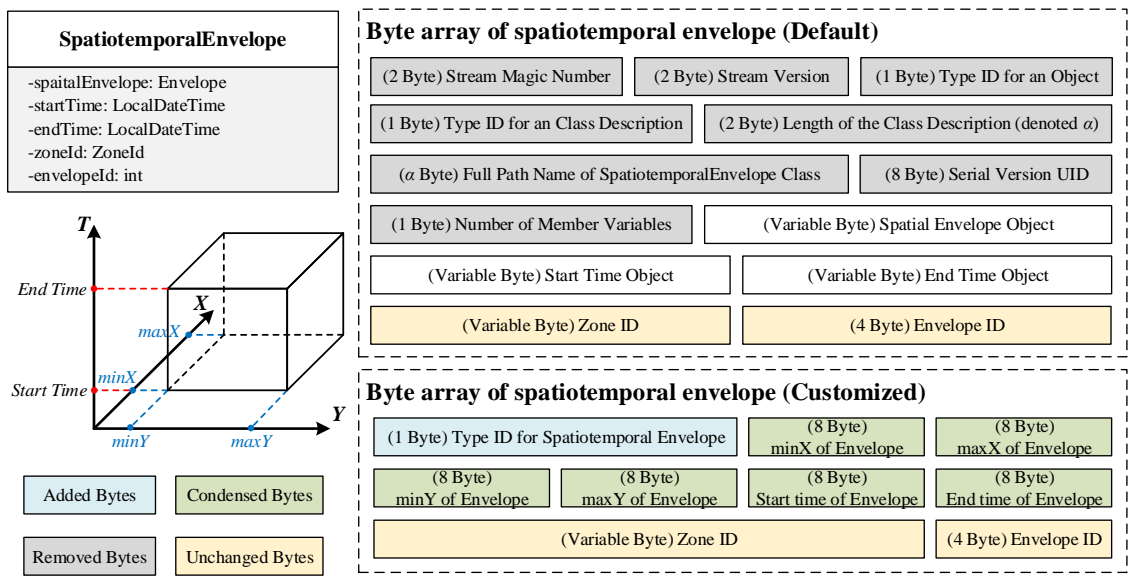

(c) Class of SpatiotemporalEnvelope and its byte array structure with default and customized serializer

**Figure 11.** Compact representation of spatiotemporal objects with customized serialization

Like spatiotemporal objects, the complete class description information of indexes is unnecessary to be transmitted between nodes for the same task. Depth-First Search (DFS) is used to traverse the tree nodes of the spatiotemporal index (Figure 12). Therefore, the serialization and deserialization of a spatiotemporal index is performed according to the structure in Figure 13. In serialization phase, right after the type ID of the spatiotemporal index, the node capacity and the spatiotemporal envelope of the entire index are transformed to byte arrays, and tree nodes are traversed in a recursive procedure. In iteration of each node, the spatiotemporal envelope of the node is serialized first, and then following by its child nodes one by one. If a node is a leaf node without any child, and the items on this node will be serialized. In deserialization phase, the same traversal strategy (DFS) is utilized. Class of the spatiotemporal index is first determined by the type ID, and its node capacity and spatiotemporal envelope are constructed. Tree nodes in the index are then constructed recursively. With these compact structures, spatiotemporal objects and indexes can be serialized / deserialized with less steps and transmitted in smaller size.

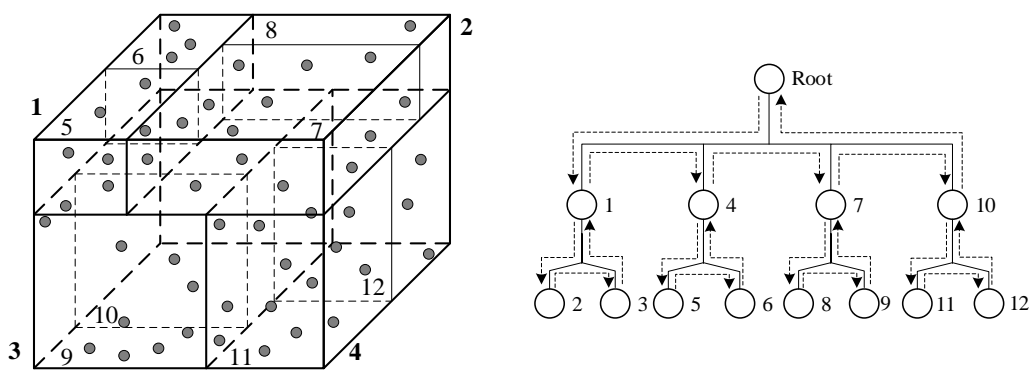

**Figure 12.** Depth-First Search (DFS) on spatiotemporal index (taking KDB-tree as an example)

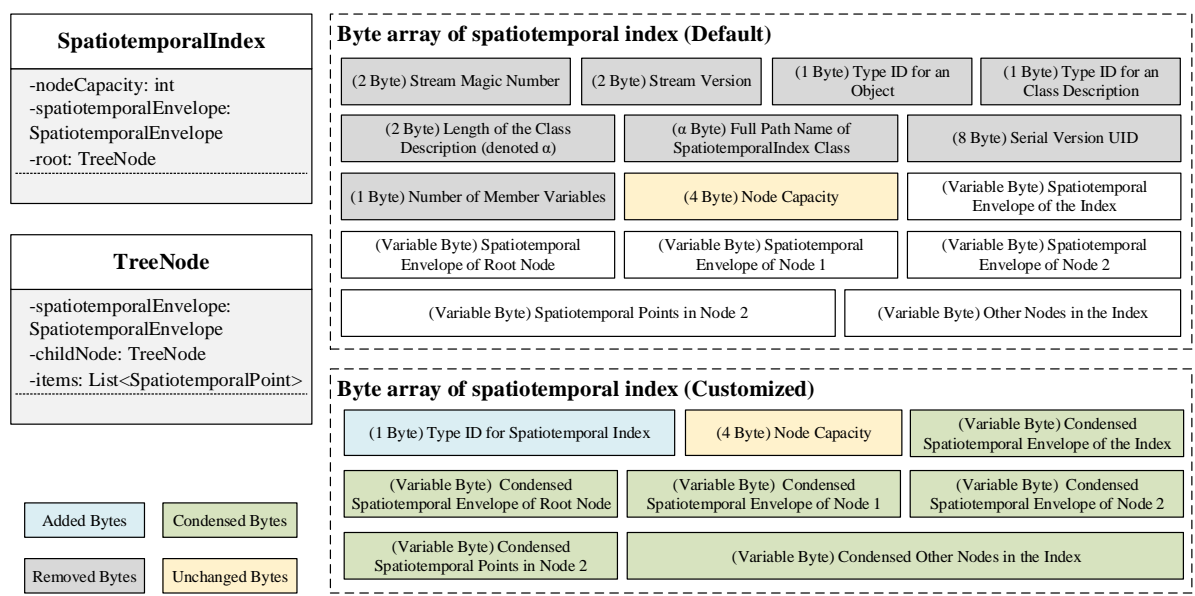

**Figure 13.** Class of spatiotemporal index and its byte array with customized serialization

## 5. Implementation

To promote the applications of space-time Ripley's K function, a multi-tier web-based visual analytics framework was designed as shown in Figure 14. The data storage tier comprises HDFS distributed file system for data management. The data processing tier is built upon Spark RDD API, spatial object API of JTS and GeoTools to support object representations and spatial operations for the distributed implementation of the space-time Ripley's K function. Abstract class *SpatiotemporalRDD* is defined to represent distributed spatiotemporal objects, and distributed spatiotemporal points and cylinders will be processed as objects of *SpatiotemporalPointRDD* and *SpatiotemporalCylinderRDD* respectively. Spatiotemporal indexes including KDB-Tree and R-Tree are built upon the spatiotemporal points as objects of *SpatiotemporalPoint* and the spatiotemporal envelopes as objects of *SpatiotemporalEnvelope*. The spatiotemporal partitioner is implemented based on KDB-Tree. Besides, the 2-tier cache for spatial and temporal weights is constructed with HashMap. On the top of data processing tier, we built web service tier and data visualization tier to make the framework more convenient to invoke the distributed space-time Ripley's K function and analyze results through a loosely-coupled and web-based an approach. The web service tier uses Jersey to provide RESTful services. In data visualization tier, the map interface is implemented upon Leaflet.js, while the base map is from OpenStreetMap, and the thematic chart is supported by D3.js.

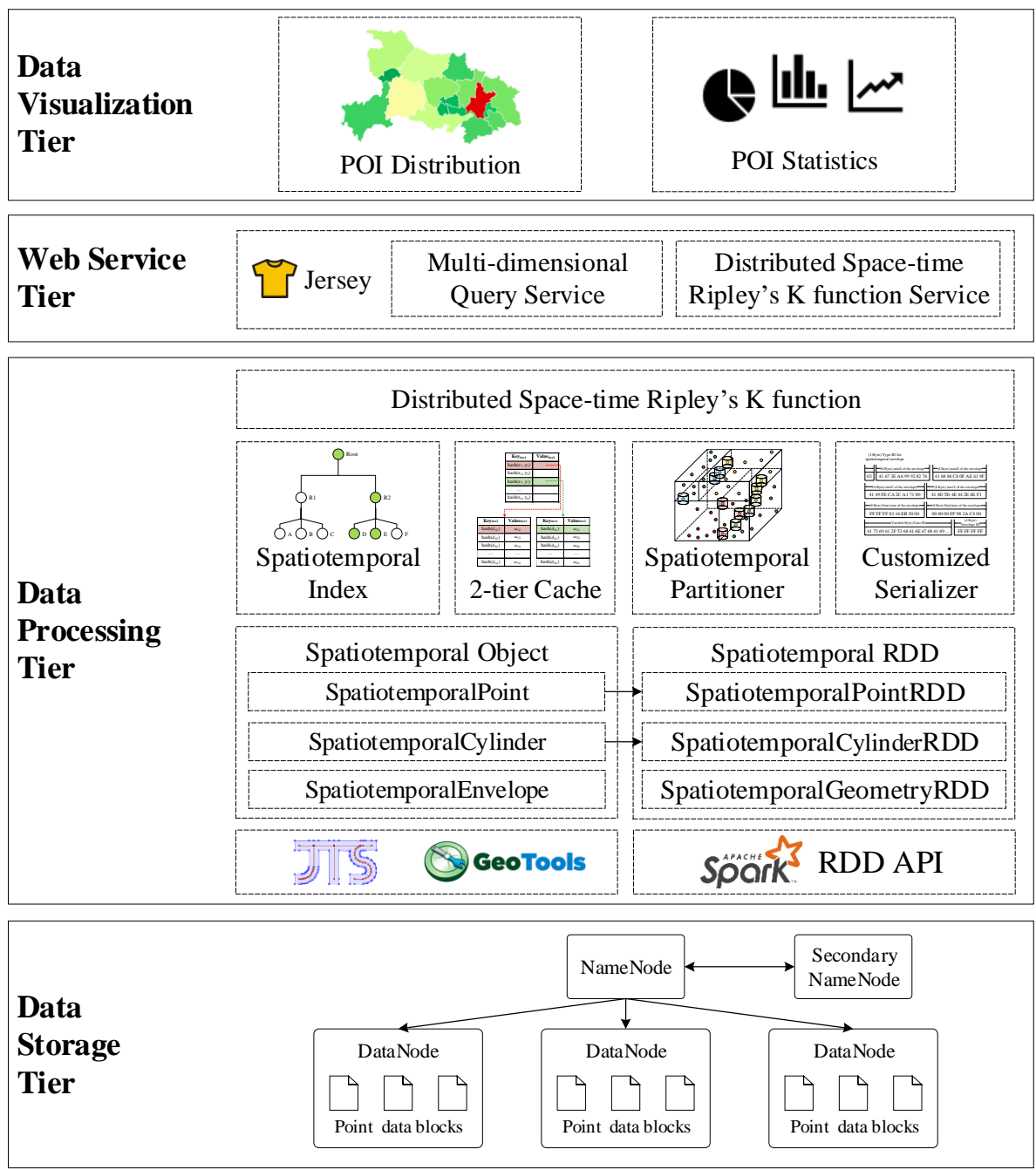

**Figure 14.** Web-based visual analytics framework for distributed space-time Ripley's K function

Based on the framework, a prototype system was developed. The system provides users with multi-dimensional filters, including multi-level spatial filter (province, city, street), multi-granular temporal filter (year, month, day), and customized categorical filter by using Nanocubes. The user can select the spatiotemporal points for analysis and configure both the space-time K function parameters and the cluster running parameters (Figure 15). For example, the user can specify study area through either MBR or the identifier of the administrative boundary, simulation method (e.g., bootstrapping and random permutation), spatiotemporal distance thresholds and step sizes. The user can also specify the total number of executors, CPU cores and memory for each executor. When the user submits the above parameters through graphical user interface (GUI), the job will be executed by the cluster. When the calculation task is completed, the respective results will be transferred in JSON format to the browser, and the user can interact with the 3D surface plots (Figure 16). For further study, the user can also download the calculation results in tabular form. All in all, through this platform, users can conduct efficient and flexible space-time point pattern analysis by using distributed computing resources without knowing the underlying technologies of the implementation.

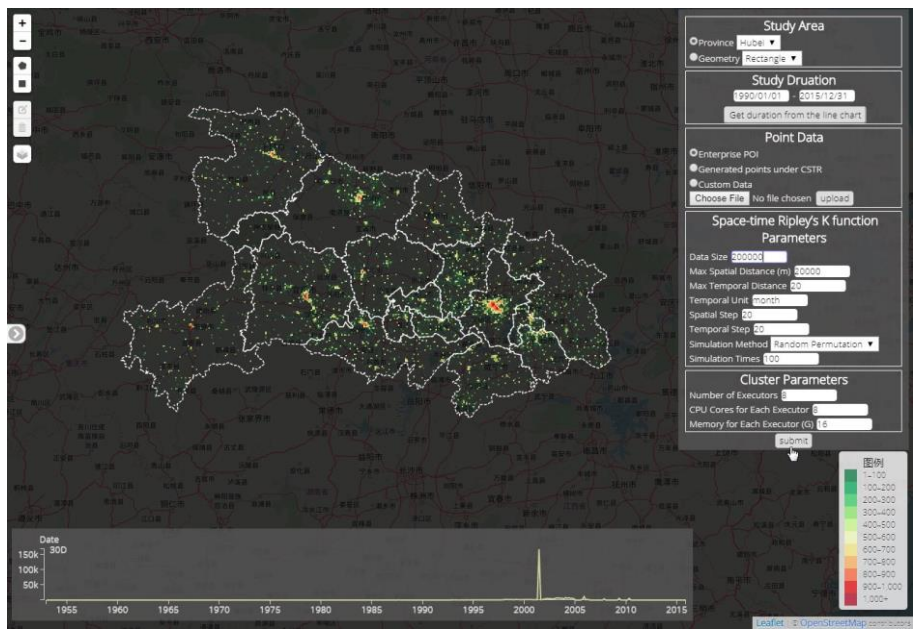

**Figure 15.** Web-based graphical user interface for parameters configuration of distributed space-time Ripley's K function

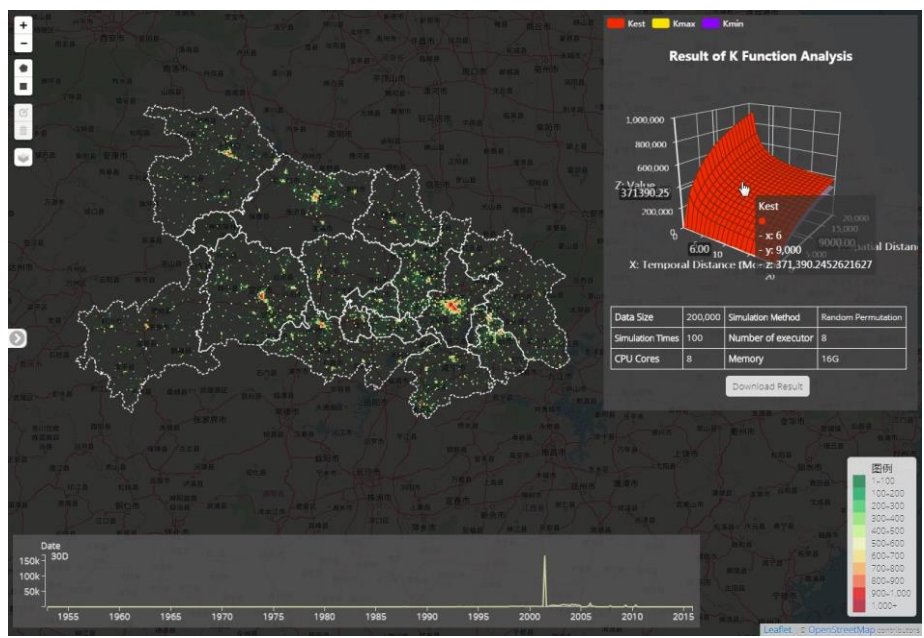

**Figure 16.** Web-based graphical user interface for interactive visualization: heatmap of spatiotemporal points and corresponding 3D surface plots for estimated and simulated results of distributed space-time Ripley's K function

## 6. Experiments and Results

### 6.1. Experiment design

Experiments were designed to evaluate the effectiveness of optimization strategies and the acceleration in the performance of the enhanced distribution implement in Spark towards the computation of the K function. The effects of the spatiotemporal index and the 2-tier cache were examined respectively in Section 6.2 and 6.3. The effects of the spatiotemporal partitioning and the customized serialization were examined in Section 6.4 and 6.5. The overall effectiveness of these optimization strategies was examined in Section 6.6. The scalability of the proposed distributed space-time Ripley's K function in cluster was analyzed in Section 6.7. A use case of optimized space-time Ripley's K function using enterprise POI data was shown in Section 6.8.

Experiments in the following sub-sections were conducted on a private cloud infrastructure supported by Apache CloudStack built upon 6 physical nodes. Each physical node that works as the agent of this private cloud has 24 CPU cores of 2.4 GHz and 64GB memory. They are connected by local area network with 1 Gbps. 9 VMs with 8 virtual CPU cores of 2 GHz and 16GB memory running CentOS 7 were created on the private cloud. 1 VM served as master, and the other 8 VMs served as workers were evenly hosted on 4 physical nodes. Experiments to examine effects of spatiotemporal index and 2-tier cache were conducted on one VM. Other experiments were conducted on entire Spark cluster.

The point dataset used in experiments was enterprises registration data in Hubei Province, China (369,826 points in total) from 1949 to 2015 recorded by the bureaus of Administration for Industry and Commerce (AIC) of China. The dataset was cleaned with imputation methods [56], and the study area was administrative boundary of Hubei Province. Spatiotemporal point analysis on these enterprise entities could support evaluation and assessment of industry concentration. To make convenience of calculation for space-time Ripley's K function, the addresses and registration dates of the enterprises were transformed and formatted, where addresses were geocoded and further transformed to projected coordinates, and registration date were converted into unified format. Experiments to examine effects of proposed strategies were conducted on subsets of this dataset, and application case of space-time Ripley's K function was presented with the complete dataset. In the following experiments, unless otherwise specified, space-time Ripley's K function was estimated and simulated from 0 to 20 km and 0 to 20 months with 1 km as spatial step and 1 month as temporal step.

In order to evaluate the performance of distributed space-time Ripley's K function, speedup factor (SF) is used to measure how much speedup of optimization strategies achieved, which is calculated as formula 5.

$$SF = T_{original}/T_{optimized} \quad (5)$$

where $T_{original}$ and $T_{optimized}$ are the execution time of the original and optimized space-time Ripley's K function respectively. Besides, acceleration factor (AF) is also defined to measure how much acceleration of the distributed system achieved:

$$AF = T_{standalone}/T_{distributed} \quad (6)$$

where $T_{standalone}$ and $T_{distirbuted}$ are the execution time of the space-time Ripley's K function performed on standalone machine and distributed cluster respectively. All the execution time mentioned above excluded time spent on data input and output, allowing for comparison of efficiency on calculation.

*6.2. Effects of spatiotemporal index*

Based on spatiotemporal index, the query scope for neighbor points that were within spatial distance $s$ and temporal distance $t$ from the center point could be narrowed quickly through comparison with spatiotemporal scope of tree node in the index (Section 4.2). The degree to which the times of inner traversals could be reduced depends on the maximum spatial distance $s_{max}$ and temporal distance $t_{max}$. A subset of the point dataset (n=50,000) introduced in Section 6.1 was used in this experiment. According to Ripley's rule of thumb [13], all $s_{max}$ selected were not greater than one quarter of the smaller side of enclosing rectangle from the study area, and all $t_{max}$ selected were not greater than one quarter of the duration of study period. In this experiment, the proportion of spatial query scope to study area and temporal query scope to study duration are both set from 1/128 to 1/4 exponentially. As shown in table 3, the index building time is much less than query and calculation time, and almost has no influence on total execution time. So, the execution time of space-time Ripley's K function with spatiotemporal index was compared to that without index to investigate the effect of spatiotemporal index.

**Table 3**. Spatiotemporal index building time under increasing query scope

| Time of index building (s) | Proportion of query scope to study area | | | | | |
|---|---|---|---|---|---|---|
| | 1/128 | 1/64 | 1/32 | 1/16 | 1/8 | 1/4 |
| Estimation | 0.2 | 0.1722 | 0.165 | 0.17167 | 0.1814 | 0.157 |
| Simulation | 0.0966 | 0.1254 | 0.092 | 0.104 | 0.138 | 0.085 |

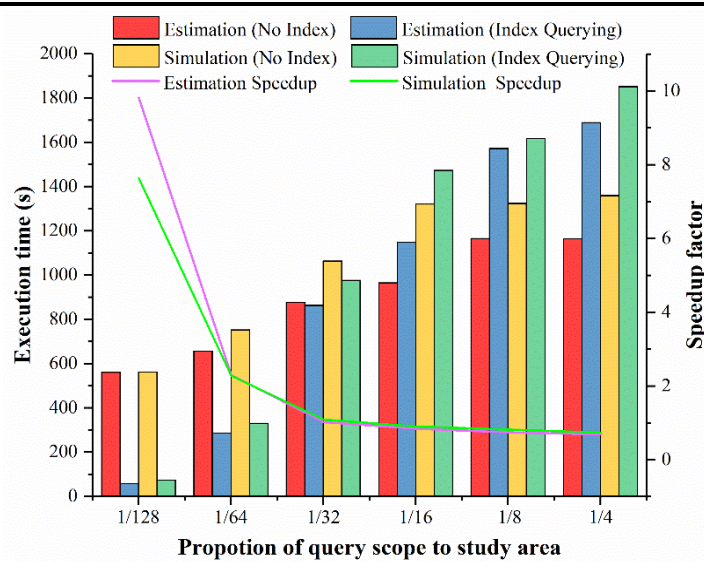

**Figure 17.** Comparison of execution times and speedup factors of space-time Ripley's K function with and without spatiotemporal index under increasing query scopes

The results in Figure 17 indicate that spatiotemporal index would increase the performance of estimation and simulations at relatively short spatial and temporal distance, while as the spatiotemporal distance increases the performance for indexing querying become less effective and even worse than the solution without indexing. The speedup factor, defined as the ratio of execution time with spatiotemporal index to that without index, was higher than 1 when $s_{max}$ and $t_{max}$ were less than 1/32 of smaller side length of spatial boundary and temporal period respectively. It could be explained that the query scopes at higher $s_{max}$ and $t_{max}$ were not partial enough, then most of the tree nodes would be traversed and the index would lose its effect. However, at lower $s_{max}$ and $t_{max}$, the speedup factor for estimation and simulation could reach 9.8 and 7.6 respectively, because the query scope would only interest with a small amount of tree nodes in the index. Meanwhile, point distribution has huge impact on the effectiveness of index. When the dataset is extremely skewed that most of points concentrated in certain peak areas, like our experimental dataset shown in figure 15, the spatiotemporal query scope will overlaps with large proportion of R-tree nodes in index, and hence weaken the filtering capability of the index seriously. Therefore, this experiment gives us suggestion on when we should use the indexing during point pair acquisition in space-time Ripley's K function implementation.

*6.3. Effects of 2-tier weight cache*

The 2-tier cache was designed to avoid repetitive calculation for spatial and temporal edge effect correction weights (Section 4.3). As the time complexity of spatiotemporal isotropic correction method is linear correlated to the complexity of boundary, this experiment was performed on the same point dataset (n=50,000) at the same maximum spatial and temporal distance but with boundaries composed of different number of vertexes. The execution time of space-time Ripley's K function with 2-tier cache was compared to that without cache to investigate the effect of 2-tier cache.

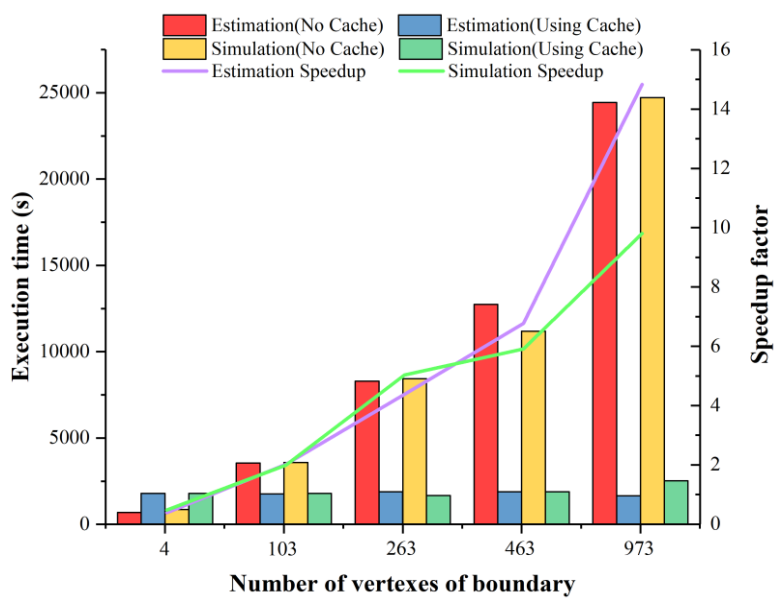

**Figure 18.** Comparison of execution times and speedup factors of space-time Ripley's K function with and without weight caching under increasing number of vertexes of study area boundary

Figure 18 shows that 2-tier cache can eliminate the influence brought by the complexity of boundary. Execution times of estimation and simulations are independent with to the boundary when using the cache, while the execution times of the counterpart solution without cache do increase linearly with the number of vertexes of the boundary. The speedup factor, defined as the ratio of execution time with 2-tier cache to that without cache, was higher than 1 when the number of vertexes on the boundary was greater than 100, and it became greater with more complex boundary. However, when the boundary is extremely simple (like minimum bounding rectangle), the calculation of weights could be quickly finished, while the overhead of 2-tier cache caused by resizing and hash collision will offset the benefits of cache. Meanwhile, we can also find that the differences between estimation and simulation was little. It might be because the edge effect of the point dataset was not significant, and most of weight was 1. If more points are located nearby boundary of study area and need edge correction, the execution time of estimation should be longer than that of simulation theoretically. Thus, 2-tier cache could decrease time cost of space-time Ripley's K function with complex boundary; but for simple boundary, there is no need to enable the 2-tier cache.

*6.4. Effects of spatiotemporal partitioning*

Through spatiotemporal partitioning, spatiotemporally neighboring points would be organized in the same partition and data redundancy would be minimized, thus accelerating the performance of space-time Ripley's K function in distributed system. Fewer partitions would result in less data redundancy in the whole system, but more memory would be required for each handle task; while more partitions would bring fine-grained tasks for better scheduling but generate more data redundancy. Hence, to maximize the acceleration by distributed system but avoid Out-Of-Memory error, spatiotemporal points should be partitioned with an appropriate number. This experiment was conducted on the point dataset (n=200,000) at the same maximum spatial and temporal distance with the same boundary using partitions of varying number. The execution time of space-time Ripley's K function with spatiotemporal partitioning was compared to that with hash partitioning, the default method in Apache Spark, to investigate the effect of spatiotemporal partitioning.

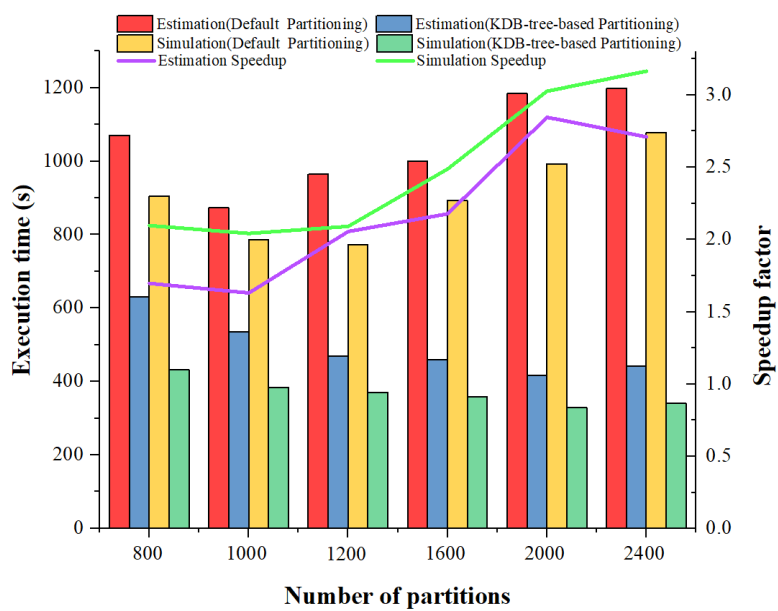

**Figure 19.** Comparison of execution times and speedup factors of space-time Ripley's K function with and without KDB-tree-based spatiotemporal partitioning under increasing number of partitions

Figure 19 illustrates that spatiotemporal partitioning could avoid unnecessary data redundancy and accelerate distributed space-time Ripley's K function. Few partitions lead to less data redundancy but bring longer average execution time, while more partitions result in fine-grained tasks for scheduling but also require more data transmission. So, the execution times of estimation and simulations under default partitioning method decreased firstly and then increased. Therefore, optimized partitioning needs to leverage execution time and transmission time by carefully adjusting number of partitions. The speedup factor, defined as the ratio of execution time with spatiotemporal partitioning to that with hash partitioning, was higher than 1 consistently and grew up with the increasing number of partitions till 2000 partitions. The optimal number of partitions for KDB-tree-based partitioning is larger than that for default partitioning, which indicated that spatiotemporal proximity could reduce invalid data transmission and improve the performance of task scheduling. Therefore, spatiotemporal partitioning could effectively improve the performance of space-time Ripley's K function in distributed system.

*6.5. Effects of customized serialization*

The customized serialization was aimed to provide compact representation of spatiotemporal objects and indexes for space-time Ripley's K function, resulting in less data transmission and faster serialization / deserialization. The volume of data transmission in distributed systems is mainly related to point data size for space-time Ripley's K function. Therefore, this experiment was conducted on different number of point datasets at the same spatial and temporal distances with the same boundary. The execution time of space-time Ripley's K function with default serialization was compare to that with customized serializer to investigate the effect of customized serializer (Figure 20).

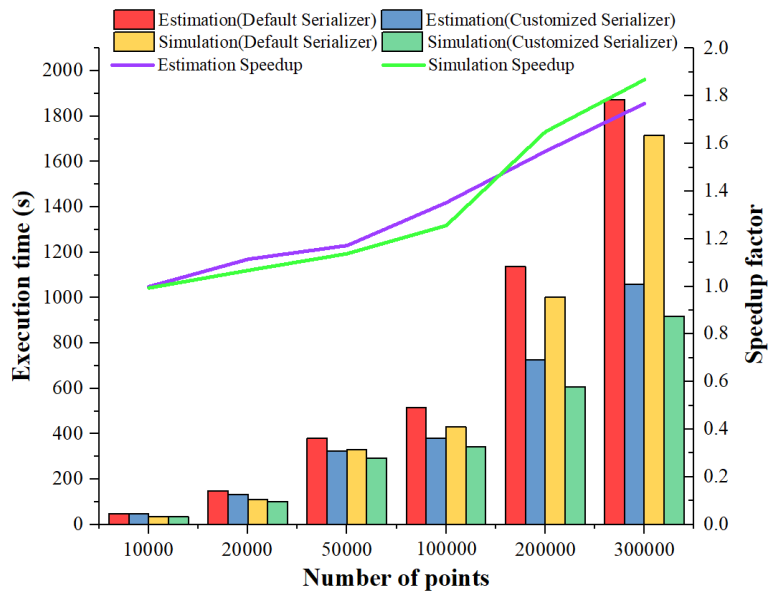

**Figure 20.** Comparison of execution times and speedup factors of space-time Ripley's K function with default serializer and customized serializer under increasing number of points

**Table 4.** Comparison of bytes sizes and serialization / deserialization times with default and customized serializer under increasing number of points

| Number of Points | $Size_D$ | $Size_C$ | CR | $Time_D^S$ | $Time_C^S$ | $SR^S$ | $Time_D^D$ | $Time_C^D$ | $SR^D$ |
|---|---|---|---|---|---|---|---|---|---|
| 10000 | 15.43 | 0.67 | 23.11 | 0.11 | 0.01 | 9.55 | 0.55 | 0.01 | 42.97 |
| 20000 | 30.86 | 1.34 | 23.11 | 0.21 | 0.02 | 11.26 | 1.17 | 0.03 | 41.61 |
| 50000 | 77.15 | 3.34 | 23.11 | 0.51 | 0.05 | 10.33 | 2.85 | 0.06 | 45.90 |
| 100000 | 154.30 | 6.68 | 23.11 | 1.14 | 0.09 | 12.30 | 5.87 | 0.13 | 45.84 |
| 200000 | 308.61 | 13.35 | 23.11 | 2.10 | 0.19 | 11.21 | 11.36 | 0.28 | 41.29 |
| 300000 | 462.91 | 20.03 | 23.11 | 2.96 | 0.28 | 10.76 | 16.59 | 0.35 | 47.12 |

Notes: unit of data volume is MB; unit of time is second. $Size_D$ is data volume of the byte arrays from default serializer; $Size_C$ is data volume of the byte arrays from customized serializer; CR is compression ratio; $Time_D^S$ is serialization time with default serializer; $Time_C^S$ is serialization time with customized serializer; $SR^S$ is speedup ratio of serialization phase; $Time_D^D$ is deserialization time with default serializer; $Time_C^D$ is deserialization time with customized serializer; $SR^D$ is speedup ratio of deserialization phase.

As is shown in Figure 20, customized serialization could reduce overhead involve with data transmission and improve performance of distributed space-time Ripley's K function. The speedup factor, defined as the ratio of execution time with default serializer to that with customized serializer, was higher than 1 consistently and grew up with the increasing number of points. More spatiotemporal points would generate more spatiotemporal cylinders and envelopes, and they would be serialized and deserialized more frequently, so customized serialization could play a more significant role for accelerating distributed space-time Ripley's K function. Table 4 compares the data volume and serialization/deserialization time cost of the default serializer and customized serializer for spatiotemporal points detailly. Since data transmission is only a portion of entire distributed calculation process, if only considering time cost of serialization and deserialization for the spatiotemporal objects, we can obtain much higher speedup ratios than that of the total execution time. Table 4 also reveals relative stable compression ratio and speedup ratio for serialization/deserialization during repeated tests. Besides, when spatiotemporal indexes were adopted to optimized space-time Ripley's K function, the customized serialization would further increase the efficiency.

*6.6. Overall effectiveness of four optimization strategies*

The overall effectiveness of the four optimization strategies was evaluated by comparing execution time of distributed space-time Ripley's K function with and without any procedure optimizations. The original algorithm and optimized algorithm are both Spark-based and tested on the same clustering computing environment with 8 worker nodes. As the complexity of computation for space-time Ripley's K function was determined by the data volume, the experiment was carried out on different size of point datasets. The maximum spatial and temporal distance thresholds were 20km and 20 months, which are 1/24 to the smaller side length of study area and 1/32 to the study duration. The number of partitions is adapted to the point data size. Customized serialization was only registered in optimized group.

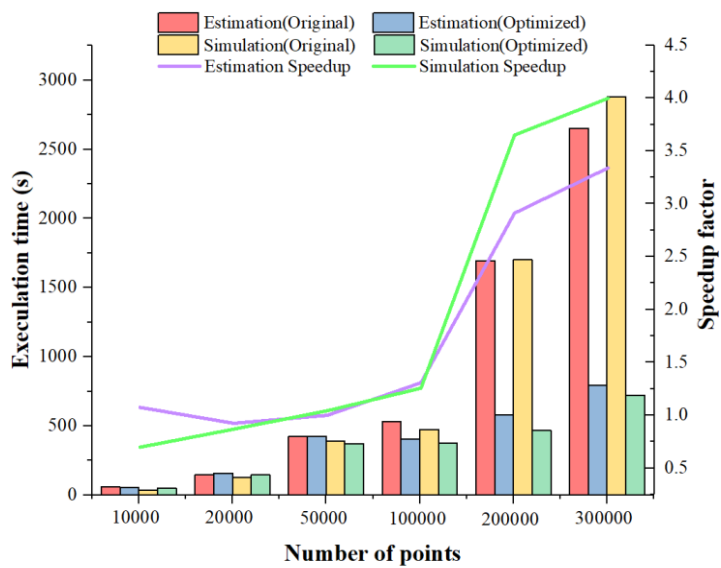

**Figure 21** Overall performance comparison of execution times and speedup factors of space-time Ripley's K function with and without distributed optimization under increasing data volume

Figure 21 demonstrates that the speedup factor achieved by four optimization strategies increases as data size increases and was higher than 1 when the data size larger than 50,000. Although the maximum spatial and temporal distance thresholds setting and extremely skewed point distribution in this experiment make the spatiotemporal index almost noneffective on time efficiency improvement as discussed in section 6.2, the estimation still achieved about 1.25 and 3.3 times speedup at data size 100,000 and 300,000 respectively. Meanwhile, the speedup for simulation was higher than that for estimation on large data sizes because of less calculation for spatiotemporal weight. Spatiotemporal index reduced unnecessary inner traversals to quickly acquire qualified point pairs under spatial and temporal thresholds. 2-tier cache enabled spatiotemporal edge correction weights reusage to avoid repetitive calculations. Spatiotemporal partitioning considered spatiotemporal distribution of the points and less data redundancy was needed to obtain correct calculation result in distributed system. Customized serialization decreased the size of serialized bytes and shorten the time cost of data transmission. As a result, the four optimized strategies effectively improve the performance of spatiotemporal point pattern analysis using space-time Ripley's K function.

*6.7. Scalability Analysis and overall speedup*

The scalability of distributed Ripley's K function was evaluated by comparing performance of estimations and simulations on different number of nodes in the cluster. As is shown in Figure 20, the effects of overall optimizations become significant when data size reaches more than 200,000. This experiment was performed on the point dataset (n=200,000) at spatial distance from 0 to 20km and temporal distance from 0 to 20 months with optimal number of partitions inferred from Figure 19. The execution time of space-time Ripley's K function on 2 to 8 nodes were compared to that on standalone computer to investigate the scalability.

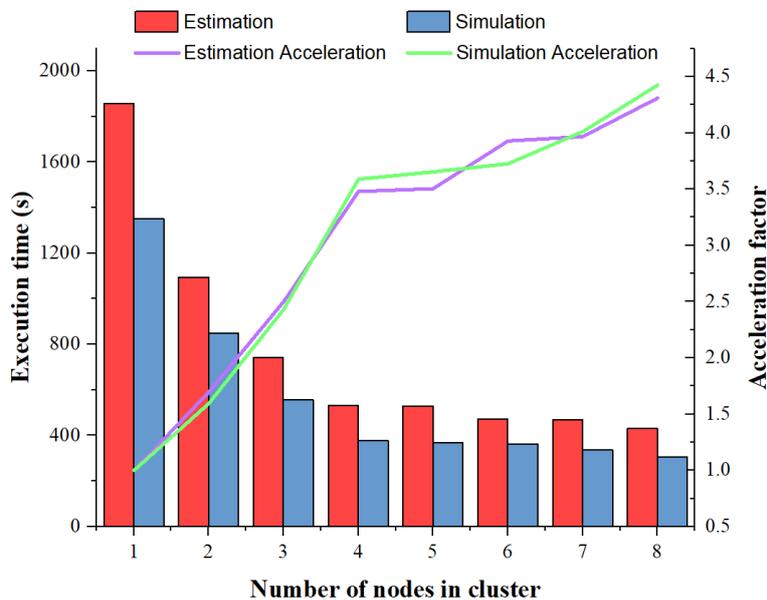

**Figure 22.** Comparison of execution times and acceleration factors of space-time Ripley's K function on increasing number of nodes in cluster

Figure 22 shows that the acceleration factor achieved varied with number of nodes in cluster. For 2 to 4 nodes, the acceleration is relatively significant. When the number of nodes is more than 5, the drop of execution time become less and less. It could be inferred that for data size like 200,000, 4 nodes with 8 CPU cores and 16G memory are sufficient to perform space-time Ripley's K function efficiently. For analysis on different data size, the required computing resources are also different.

Table 5. Overall speedup by adopting optimization strategies and clustering computing

| Number of Points | Execution time (s) | | | | Speedup | |
|---|---|---|---|---|---|---|
| | Original on 1-node | | Optimized on 8-nodes | | | |
| | Estimation | Simulation | Estimation | Simulation | Estimation | Simulation |
| 10,000 | 120.769 | 104.459 | 54.907 | 50.831 | 2.20 | 2.06 |
| 20,000 | 584.731 | 517.169 | 158.861 | 146.507 | 3.68 | 3.53 |
| 50,000 | 1891.986 | 1672.767 | 424.368 | 372.554 | 4.46 | 4.49 |
| 100,000 | 2136.239 | 1992.250 | 407.021 | 378.037 | 5.25 | 5.27 |
| 200,000 | 4313.731 | 4296.120 | 582.150 | 466.303 | 7.41 | 9.21 |
| 300,000 | 6427.267 | 6683.379 | 793.656 | 720.325 | 8.10 | 9.28 |

Notes: "Original on 1-node" refer to the original non-optimized algorithm ran on single VM using Spark local mode, while "Optimized on 8-nodes" is for the optimized algorithm ran on Spark cluster with 8 worker nodes.

Table 5 shows overall speedup by adopting the proposed optimization strategies and Apache Spark clusters comparing with the original algorithm ran on single VM with Spark local mode. We can find that the overall speedup increases as the data size increases, which can achieve around 8 times for estimation and 9 times for simulation at data size 300,000. Nevertheless, the speedup is still limited by the restricted network speed (1 Gbps), hypervisor overhead, partition size setting, and can be improved furtherly.

The experiment results from Section 6.2 to 6.7 could provide guidelines for potential applications of distributed space-time Ripley's K function. The effects of the optimization strategies are influenced by point distribution and parameter settings of space-time Ripley's K function, including data size, spatiotemporal distance thresholds and complexity of boundary. Table *A* in appendix demonstrates a significant speedup ratio gained on a synthetic dataset. In general, in leveraging the time efficiency and computing resource cost, the distributed method should be adopted with large data size. Meanwhile, the cluster size should be adjusted along with the data size, e.g., 4 nodes for 200,000 points in this experiment. The recommended memory for each node is at least 16GB to avoid Out-Of-Memory (OOM) Problem in distributed environment. When spatiotemporal distance thresholds exceed certain range, e.g., 1/16 length of smaller side of MBR of study area and 1/16 length of study duration, spatiotemporal index for point pair acquisition might introduce extra execution time and could be disabled. Besides, if the geometry boundary of the study area is oversimplified with limited vertexes just like the MBR of study area, 2-tier cache could be switch off to avoid its negative effects. As for spatiotemporal partitioning and customized serialization, they could be enabled by default since the experiments have shown their advantages comparing with default solutions, but further investigation is still needed.

*6.8. Application case*

With the developed distributed space-time Ripley's K function, spatiotemporal point pattern analysis could be conducted smoothly for large datasets. The distributed space-time Ripley's K function was evaluated using the real-world dataset, the entire enterprise POI in Hubei Province, China (369,826 points, 100 simulations). Compared to the original algorithm that took 53 hours, the optimized algorithm implemented in the same distributed environment with 8 worker nodes could complete the task 4 times faster. To learn the industrial aggregation effect in some area, one might need to examine the spatiotemporal pattern across different scales. In this example, through multiscale point pattern analysis we can compare the estimated $\hat{L}(s,t)$ with the simulated $\hat{L}(s,t)$ across different spatial and temporal distances and then draw conclusion about the spatiotemporal clustering scope of the enterprises. From Figure 23, we can reject null hypothesis of CSTR for enterprise POI in Hubei Province at 0.01 level of significance, and spatial distance within 3 km, temporal distance within 1.5 month, are the most significant spatiotemporal clustering range for these point events.

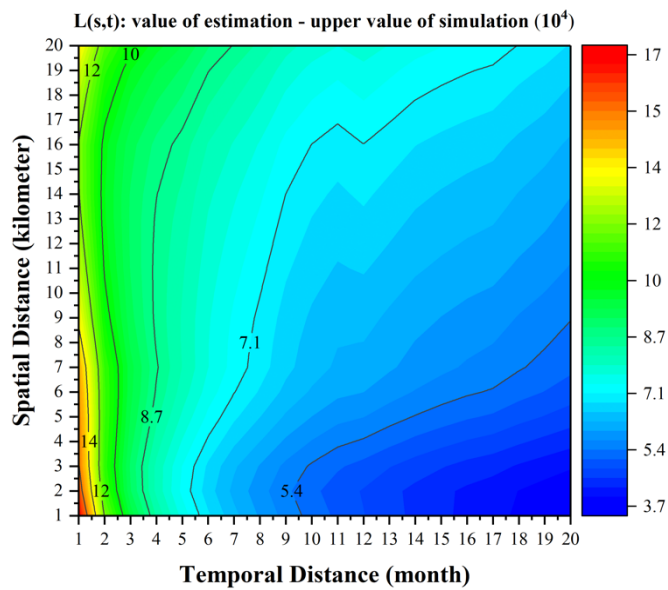

**Figure 23.** Absolute difference between L value of estimation and upper L value of simulations

**7. Conclusions and Future Work**

This paper proposed a performance-improved distributed space-time Ripley's K function for better supporting spatiotemporal point pattern analysis. In particular, spatiotemporal-index-based point pair acquisition, 2-tier-cache weight reuse, spatiotemporal partitioning and customized serialization for spatiotemporal objects and indexes were designed to improve the procedure space-time Ripley's K function. Based on these optimization strategies, we implemented space-time Ripley's K function by using Apache Spark. Theoretical analysis and experiment evaluations highlight that the proposed method can reduce the time complexity of space-time Ripley's K function for large datasets. Performance evaluations revealed that the first two strategies could speed up the calculation with appropriated query scope setting even in a stand-alone execution environment by optimizing the calculation procedure of space-time Ripley's K function; while the last two strategies accelerate the distributed workflow furtherly by balancing workloads and decreasing data transmission overheads. By integrating these optimization methods into a distributed computing framework, Space-time Ripley's K function could be applied to big data scenarios with less time and economic cost.

Beside Apache Spark, the proposed optimization method is generic and could be adopted to other distributed computing environments (DCEs) with moderate modification, e.g., Hadoop, Flink. Although, most of the optimizations are implemented upon Spark in-memory architecture and RDD, the foundational mechanisms of caching, indexing and partitioning and serialization are well-supported by many DCEs. Developers and researchers can use the idea to implement similar optimization by using spatial extensions in those DCEs. Meanwhile, the major optimization is implemented upon fundamental spatial libraries such JTS and GeoTools in Java, and could be portable to any platform using Java Virtual Machine (JVM).

Using space-time Ripley's K function as an example, this study demonstrated how spatial principles and distributed frameworks could be leveraged to optimize and accelerate spatial data processing. Spatial principles indicate the spatial and temporal characteristics among the events and phenomenon and have been applied for computational optimization in earth sciences [57,58], physical science [59,60] and so on. In many geocomputation cases, spatial closer events has more significant impact than that on far way events according to first law of geography [2], which involves domain decomposition [23,42] and explains the role of spatiotemporal indexes for eliminating unnecessary computing. Similar spatiotemporal characteristics of phenomenon would result in similar statistical values, especially for space-time distance and weight calculation in many other spatial analysis methods, e.g., Kernel Density Estimation (KDE) and Geographical Weighted Regression (GWR), so repetitive calculation could be avoided by cache mechanism. Distributed frameworks make it easier to reach superior performance on commodity computers, and the user-friendly APIs of mainstream distributed frameworks lower the learning curve for developing reliable parallel algorithms. This study makes contributions on how to lower the barrier to time-consuming spatial analysis methods and promote their broader applications for large datasets.

Future work would focus on the following directions. Firstly, adaptive control and adoption guideline for the proposed optimization strategies could be further investigated, e.g., applicable scenarios detection for spatiotemporal index and 2-tier cache, optimal parameter selection for spatiotemporal partitioning. Secondly, this study concentrated on univariate homogeneous isotropic spatiotemporal point analysis using space-time Ripley's K function. Distributed Ripley's K function could be further extended to bivariate [61], inhomogeneous [62], anisotropic [63] point pattern analysis or combinations of them, and better understanding could be facilitated on complex point process in the real world. Thirdly, distributed frameworks are still evolving, the acceleration of distributed Ripley's K function on the combination of various computing resources could be evaluated [64,65], such as many-core GPUs, Field-Programmable Gate Array (FPGA) [66].

**Acknowledgement:** This paper is supported by National Key R&D Program of China (No. 2017YFB0503704 and No. 2018YFC0809806) and National Natural Science Foundation of China (No. 41971349, No. 41501434 and No. 41371372). Thanks to the anonymous reviewers for their valuable suggestions.

**Appendix**

**Table A.** Overall speedup on a synthetic dataset by adopting optimization strategies and clustering computing

| Number of Points | Execution time (s) | | | | Speedup | |
|---|---|---|---|---|---|---|
| | Original on 1-node | | Optimized on 8-nodes | | | |
| | Estimation | Simulation | Estimation | Simulation | Estimation | Simulation |
| 100,000 | 1126.132 | 1028.697 | 18.800 | 18.056 | 59.90 | 56.97 |
| 200,000 | 2292.138 | 2187.563 | 20.615 | 19.396 | 118.18 | 112.78 |
| 300,000 | 5430.595 | 5120.041 | 21.140 | 20.435 | 256.88 | 250.55 |
| 400,000 | 27582.707 | 31006.894 | 23.389 | 26.699 | 1179.30 | 1161.35 |
| 500,000 | 37266.782 | 39246.228 | 31.517 | 33.748 | 1182.43 | 1162.92 |

Notes: The point datasets are generated by ArcGIS that random distributed in United States. A simplified administration boundary of United States with 85 points is used as the boundary of the study area. The time attributes of the points are random assigned from 1949/01/01 to 2015/01/01. Space-time Ripley's K function was estimated and simulated from 0 to 10 km and 0 to 10 months as maximum spatial and time distance thresholds respectively, with 1 km as spatial step and 1 month as temporal step. We can find that the execution times of the optimized algorithm ran on 8 worker nodes grown slowly as the increase of data size and didn't show the exponential increase as that of the original non-optimized algorithm on single VM because of the random spatiotemporal data distribution. Since adjacent point pair grows linearly, the index and partitioning can work effectively, and eventually gained significant speedup ratios.

**References**

[1] R. Brian David, Statistical Inference for Spatial Processes, Cambridge: Cambridge University Press, 1988.

[2] F. A. Stewart, C. Brunsdon, M. Chalrton, Quantitative geography: perspectives on spatial data analysis, Sage, 2000.

[3] A. Hohl, M. Zheng, W. Tang, E. Delmelle, I. Casas, Spatiotemporal point pattern analysis using Ripley's K function, in: Geospatial Data Sci. Tech. Appl., 2017: pp. 155–175. doi:10.1201/b22052.

[4] K. Yuan, X. Chen, Z. Gui, F. Li, H. Wu, A quad-tree-based fast and adaptive Kernel Density Estimation algorithm for heat-map generation, Int. J. Geogr. Inf. Sci. (2019) 1–22. doi:10.1080/13658816.2018.1555831.

[5] K.E. Hendricks, M. Christman, P.D. Roberts, Spatial and Temporal Patterns of Commercial Citrus Trees Affected by Phyllosticta citricarpa in Florida, Sci. Rep. 7 (2017) 1641. doi:10.1038/s41598-017-01901-2.

[6] R. Winter-Livneh, T. Svoray, I. Gilead, Settlement patterns, social complexity and agricultural strategies during the Chalcolithic period in the Northern Negev, Israel, J. Archaeol. Sci. 37 (2010) 284–294. doi:10.1016/j.jas.2009.09.039.

[7] A. Hohl, E. Delmelle, W. Tang, I. Casas, Accelerating the discovery of space-time patterns of infectious diseases using parallel computing, Spat. Spatiotemporal. Epidemiol. 19 (2016) 10–20. doi:10.1016/j.sste.2016.05.002.

[8] K. Pandit, E. Bevilacqua, G. Mountrakis, R.W. Malmsheimer, Spatial Analysis of Forest Crimes in Mark Twain National Forest, Missouri, J. Geospatial Appl. Nat. Resour. 1 (2016) 3.

[9] G. Zhu, Y. Ge, H. Wang, A modified Ripley's K function to detecting spatial pattern of urban system, in: 2013 21st Int. Conf. Geoinformatics. IEEE, IEEE, 2013: pp. 1–5. doi:10.1109/Geoinformatics.2013.6626127.

[10] J.Y. Fu, C.F. Jing, M.Y. Du, Y.L. Fu, P.P. Dai, Study on adaptive parameter determination of cluster analysis in urban management cases, Int. Arch. Photogramm. Remote Sens. Spat. Inf. Sci. 42 (2017). doi:10.5194/isprs-archives-XLII-2-W7-1143-2017.

[11] R. Kosfeld, H.F. Eckey, J. Lauridsen, Spatial point pattern analysis and industry concentration, Ann. Reg. Sci. 47 (2011) 311–328. doi:10.1007/s00168-010-0385-5.

[12] S. Tian, J. Wang, Z. Gui, H. Wu, Y. Wang, A case study: Exploring industrial agglomeration of manufacturing industries in Shanghai using duranton and overman's K-density function, in: Int. Arch. Photogramm. Remote Sens. Spat. Inf. Sci., 2017. doi:10.5194/isprs-archives-XLII-2-W7-149-2017.

[13]   Y. Chen, K. Qin, Z. Gui, H. Wu, Exploring spatial agglomeration of China's secondary industry based on registration data of industrial and commercial enterprises, J. Liaoning Tech. Univ. (Natural Sci. 37 (2018) 602–610.

[14]   J. Sporring, R. Waagepetersen, S. Sommer, Generalizations of Ripley's K-function with Application to Space Curves, in: Int. Conf. Inf. Process. Med. Imaging, 2019: pp. 731–742. doi:10.1007/978-3-030-20351-1.

[15]   C. Yang, M. Sun, K. Liu, Q. Huang, Z. Li, Z. Gui, Y. Jiang, J. Xia, M. Yu, C. Xu, P. Lostritto, N. Zhou, Contemporary computing technologies for processing big spatiotemporal data, in: Space-Time Integr. Geogr. GIScience, 2015: pp. 327–351. doi:10.1007/978-94-017-9205-9.

[16]   M.F. Goodchild, Citizens as sensors: The world of volunteered geography, GeoJournal. 69 (2007) 211–221. doi:10.1007/s10708-007-9111-y.

[17]   A. Baddeley, R. Turner, Spatstat: an R package for analyzing spatial point patterns, J. Stat. Softw. 12 (2005) 1–42. doi:10.18637/jss.v012.i06.

[18]   B. Rowlingson, P.J. Diggle, R. Bivand, G. Petris, S.J. Eglen, Splancs: spatial and space-time point pattern analysis, (2013).

[19]   E. Gabriel, B. Rowlingson, P.J. Diggle, stpp: An R Package for Plotting, Simulating and Analyzing Spatio-Temporal Point Patterns, J. Stat. Softw. 53 (2013) 1–29.

[20]   K. Hu, Z. Gui, X. Cheng, H. Wu, S. McClure, The Concept and Technologies of Quality of Geographic Information Service: Improving User Experience of GIServices in a Distributed Computing Environment, ISPRS Int. J. Geo-Information. 8 (2019) 118. doi:10.3390/ijgi8030118.

[21]   Q. Guan, P.C. Kyriakidis, M.F. Goodchild, A parallel computing approach to fast geostatistical areal interpolation, Int. J. Geogr. Inf. Sci. 25 (2011) 1241–1267.

[22]   G. Zhang, Q. Huang, A.X. Zhu, J.H. Keel, Enabling point pattern analysis on spatial big data using cloud computing: optimizing and accelerating Ripley's K function, Int. J. Geogr. Inf. Sci. 30 (2016) 2230–2252. doi:10.1080/13658816.2016.1170836.

[23]   W. Tang, W. Feng, M. Jia, Massively parallel spatial point pattern analysis: Ripley's K function accelerated using graphics processing units, Int. J. Geogr. Inf. Sci. 29 (2015) 412–439. doi:10.1080/13658816.2014.976569.

[24]   G. Manogaran, D. Lopez, N. Chilamkurti, In-Mapper combiner based MapReduce algorithm for processing of big climate data, Futur. Gener. Comput. Syst. 86 (2018) 433–445. doi:10.1016/j.future.2018.02.048.

[25]   T. Nguyen, M. Larsen, B. O'Dea, H. Nguyen, D.T. Nguyen, J. Yearwood, D. Phung, S. Venkatesh, H. Christensen, Using spatiotemporal distribution of geocoded Twitter data to predict US county-level health indices, Futur. Gener. Comput. Syst. (2018). doi:10.1016/j.future.2018.01.014.

[26]   A. Cano, A survey on graphic processing unit computing for large-scale data mining, WIREs Data Min. Knowl Discov. e1232 (2017). doi:10.1002/widm.1232.

[27]   H.R. Asaadi, D. Khaldi, B. Chapman, A comparative survey of the HPC and big data paradigms: Analysis and experiments, Proc. - IEEE Int. Conf. Clust. Comput. ICCC. (2016) 423–432. doi:10.1109/CLUSTER.2016.21.

[28]   J. Lu, R.H. Güting, Parallel SECONDO: Boosting database engines with Hadoop, in: 2012 IEEE 18th Int. Conf. Parallel Distrib. Syst., 2012: pp. 738–743. doi:10.1109/ICPADS.2012.119.

[29]   A. Aji, F. Wang, H. Vo, R. Lee, Q. Liu, X. Zhang, J. Saltz, Hadoop-GIS: A High Performance Spatial Data Warehousing System Over MapReduce, Proc. VLDB Endow. 6 (2013) 1009–1020.

[30]   A. Eldawy, M.F. Mokbel, SpatialHadoop: A MapReduce Framework for Spatial Data, in: 2015 IEEE 31st Int. Conf. Data Eng., 2015: pp. 1352–1363.

[31]   S. You, J. Zhang, L. Gruenwald, Large-scale spatial join query processing in Cloud, in: 2015 31st IEEE Int. Conf. Data Eng. Work., 2015: pp. 34–41. doi:10.1109/ICDEW.2015.7129541.

[32]   J.N. Hughes, A. Annex, C.N. Eichelberger, A. Fox, A. Hulbert, M. Ronquest, GeoMesa: a distributed architecture for spatio-temporal fusion, Geospatial Informatics, Fusion, Motion Video Anal. V. 9473 (2015) 94730F. doi:10.1117/12.2177233.

[33]   R. Sriharsha, Magellan: geospatial analytics on spark, Https://Hortonworks.Com/Blog/Magellan-Geospatial-Analytics-in-Spark/. (2015).

[34]   D. Xie, F. Li, B. Yao, G. Li, L. Zhou, M. Guo, Simba: Efficient In-Memory Spatial Analytics, in: Proc. 2016 Int. Conf. Manag. Data. ACM, 2016: pp. 1071–1085. doi:10.1145/2882903.2915237.

[35]   J. Yu, Z. Zhang, M. Sarwat, Spatial data management in apache spark: the GeoSpark perspective and beyond, Geoinformatica. (2018) 1–42. doi:10.1007/s10707-018-0330-9.

[36]   J. Gonzalez-lopez, S. Ventura, A. Cano, Distributed Nearest Neighbor Classification for Large-Scale Multi-label Data on Spark, Futur. Gener. Comput. Syst. 87 (2018) 66–82. doi:10.1016/j.future.2018.04.094.

[37]   T. Nakaya, K. Yano, Visualising crime clusters in a space-time cube: An exploratory data-analysis approach using space-time kernel density estimation and scan statistics, Trans. GIS. 14 (2010) 223–239. doi:10.1111/j.1467-9671.2010.01194.x.

[38]   E. Gabriel, Estimating Second-Order Characteristics of Inhomogeneous Spatio-Temporal Point Processes, Methodol. Comput. Appl. Probab. 16 (2014) 411–431. doi:10.1007/s11009-013-9358-3.

[39]   P.J. Diggle, A.G. Chetwynd, R. Häggkvist, S.E. Morris, Second-order analysis of space-time clustering, Stat. Methods Med. Res. 4 (1995) 124–136. doi:10.1177/096228029500400203.

[40]   P.J. Diggle, Statistical analysis of spatial and spatio-temporal point patterns, 2013.

[41]   D. Meagher, Geometric Modeling Using Octree Encoding, Comput. Graph. Image Process. 19 (1982) 129–147. http://fab.cba.mit.edu/classes/S62.12/docs/Meagher_octree.pdf.

[42]   A. Hohl, E.M. Delmelle, W. Tang, Spatiotemporal Domain Decomposition for Massive Parallel Computation of Space-Time Kernel Density, ISPRS Ann. Photogramm. Remote Sens. Spat. Inf. Sci. 2 (2015) 7–11. doi:10.5194/isprsannals-II-4-W2-7-2015.

[43]   A. Guttman, R-trees: a dynamic index structure for spatial searching, in: ACM SIGMOD Rec., 1984: pp. 47–57.

[44]   X. Xu, J. Han, W. Lu, RT-tree: An improved R-tree index structure for spatio-temporal database, in: Int. Symp. Spat. Data Handl., 1990: pp. 1040--1049.

[45]   Y. Theodoridis, M. Vazirgiannis, T. Sellis, Spatio-temporal indexing for large multimedia applications, in: Proc. Third IEEE Int. Conf. Multimed. Comput. Syst. IEEE., 1996: pp. 441–448. doi:10.1109/mmcs.1996.535011.

[46]   D. Pfoser, C.S. Jensen, Y. Theodoridis, Novel Approaches to the Indexing of Moving Object Trajectories, VLDB. (2000) 395–406. http://www.researchgate.net/publication/2395927_Novel_Approaches_to_the_Indexing_of_Moving_Object_Trajectories/file/9c960516f90ac7862e.pdf.

[47]   B.C. Giao, D.T. Anh, Improving sort-tile-recusive algorithm for R-tree packing in indexing time series, in: 2015 IEEE RIVF Int. Conf. Comput. Commun. Technol. Innov. Vis. Futur., IEEE, 2015: pp. 117–122. doi:10.1109/RIVF.2015.7049885.

[48]   S. V. Limkar, R.K. Jha, A novel method for parallel indexing of real time geospatial big data generated by IoT devices, Futur. Gener. Comput. Syst. 97 (2019) 433–452. doi:10.1016/j.future.2018.09.061.

[49]   J.L. Bentley, Multidimensional binary search trees used for associative searching, Commun. ACM. 18 (1975) 509–517. doi:10.1145/361002.361007.

[50]   X. Guan, C. Bo, Z. Li, Y. Yu, ST-hash: An efficient spatiotemporal index for massive trajectory data in a NoSQL database, in: 2017 25th Int. Conf. Geoinformatics. IEEE, 2017: pp. 1–7. doi:10.1109/GEOINFORMATICS.2017.8090927.

[51]   S.T. Leutenegger, M.A. Lopez, J. Edgington, STR: A simple and efficient algorithm for R-tree packing, in: Proc. 13th Int. Conf. Data Eng. IEEE, 1997: pp. 497–506.

[52]   T.H. Cormen, C.E. Leiserson, R.L. Rivest, C. Stein, Introduction to Algorithms, 2009. doi:10.1163/9789004256064_hao_introduction.

[53]   J.T. Robinson, The KDB-tree: a search structure for large multidimensional dynamic indexes, in: Proc. 1981 ACM SIGMOD Int. Conf. Manag. Data. ACM, 1981: pp. 10–18. doi:10.1145/582318.582321.

[54]   A. Eldawy, L. Alarabi, M.F. Mokbel, Spatial partitioning techniques in SpatialHadoop, Proc. VLDB Endow. 8 (2015) 1602–1605. doi:10.14778/2824032.2824057.

[55]   L. Opyrchal, A. Prakash, Efficient Object Serialization in Java, in: Proceedings. 19th IEEE Int. Conf. Distrib. Comput. Syst. Work. Electron. Commer. Web-Based Appl. Middlew., 1999: pp. 96–101.

[56]   F. Li, Z. Gui, H. Wu, J. Gong, Y. Wang, S. Tian, J. Zhang, Big enterprise registration data imputation: Supporting spatiotemporal analysis of industries in China, Comput. Environ. Urban Syst. 70 (2018) 9–23. doi:10.1016/j.compenvurbsys.2018.01.010.

[57]   C. Yount, A. Duran, J. Tobin, Multi-level spatial and temporal tiling for efficient HPC stencil computation on many-core processors with large shared caches, Futur. Gener. Comput. Syst. 92 (2019) 903–919. doi:10.1016/j.future.2017.10.041.

[58]   J. Xia, C. Yang, Q. Li, Using spatiotemporal patterns to optimize Earth Observation Big Data access: Novel approaches of indexing, service modeling and cloud computing, Comput. Environ. Urban Syst. 72 (2018) 191–203. doi:10.1016/j.compenvurbsys.2018.06.010.

[59]   C. Yang, H. Wu, Q. Huang, Z. Li, J. Li, Using spatial principles to optimize distributed computing for enabling the physical science discoveries, Proc. Natl. Acad. Sci. 108 (2011) 5498–5503. doi:10.1073/pnas.0909315108.

[60]   Z. Gui, M. Yu, C. Yang, Y. Jiang, S. Chen, J. Xia, Q. Huang, K. Liu, Z. Li, M.A. Hassan, B. Jin, Developing subdomain allocation algorithms based on spatial and communicational constraints to accelerate dust storm simulation, PLoS One. 11 (2016) e0152250. doi:10.1371/journal.pone.0152250.

[61]   B.D. Ripley, Modelling spatial patterns, J. R. Stat. Soc. Ser. B. 39 (1977) 172–212.

[62]   A.J. Baddeley, J. Møller, R. Waagepetersen, Non- and semi-parametric estimation of interaction in inhomogeneous point patterns, Stat. Neerl. 54 (2000) 329–350. doi:10.1111/1467-9574.00144.

[63]   J. Møller, H. Toftaker, Geometric Anisotropic Spatial Point Pattern Analysis and Cox Processes, Scand. J. Stat. 41 (2014) 414–435. doi:10.1111/sjos.12041.

[64]   P. Li, Y. Luo, N. Zhang, Y. Cao, HeteroSpark: A heterogeneous CPU/GPU Spark platform for machine learning algorithms, in: 2015 IEEE Int. Conf. Networking, Archit. Storage (NAS). IEEE, 2015: pp. 347–348. doi:10.1109/NAS.2015.7255222.

[65]   E. Ghasemi, P. Chow, Accelerating Apache Spark with FPGAs, Concurr. Comput. Pract. Exp. 31 (2019) e4222. doi:10.1002/cpe.4222.

[66]   H.F.-W. Sadrozinski, J. Wu, Applications of field-programmable gate arrays in scientific research, 2016.